\documentclass[pre,twocolumn,english,superscriptaddress,floatfix]{revtex4}

\usepackage{amsmath}  
\usepackage{amsfonts} 
\usepackage{graphicx} 
\usepackage{epsfig}
\usepackage{appendix}
\usepackage{url}

\begin{document}

\title{Elasticity of spheres with buckled surfaces}

\author{Yingzhen Tian}
\affiliation{Department of Physics, Yale University, New Haven, Connecticut}
\author{Megan McCarthy}
\affiliation{Department of Cell Biology, Yale School of Medicine, New Haven, Connecticut}
\author{Megan King}
\affiliation{Department of Cell Biology, Yale School of Medicine, New Haven, Connecticut}
\affiliation{Department of Molecular,
Cellular, and Developmental Biology, Yale University, New Haven, Connecticut}
\author{S. G. J. Mochrie}
\affiliation{Department of Physics, Yale University, New Haven, Connecticut}
\affiliation{Department of Applied Physics, Yale University, New Haven, Connecticut}

\date{\today}

\begin{abstract}
The buckling instabilities of core-shell systems, comprising an interior elastic sphere, attached to an exterior shell, have been proposed to underlie myriad biological morphologies. To fully discuss such systems, however, it is important to properly understand the elasticity of the spherical core. Here, by exploiting well-known properties of the solid harmonics, we present a simple, direct method for solving the linear elastic problem of spheres and spherical voids with surface deformations, described by a real spherical harmonic. We calculate the corresponding  bulk elastic energies, providing closed-form expressions for any values of the spherical harmonic degree ($l$), Poisson ratio, and shear modulus. We find that the elastic energies are independent of the spherical harmonic index ($m$). Using these results, we revisit the buckling instability experienced by a core-shell system comprising an elastic sphere, attached within a membrane of fixed area, that occurs when the area of the membrane sufficiently exceeds the area of the unstrained sphere [C. Fogle, A. C. Rowat, A. J. Levine and J. Rudnick, {\em Phys. Rev. E} {\bf 88}, 052404 (2013)]. We determine the phase diagram of the core-shell sphere's shape, specifying what value of $l$ is realized as a function of the area mismatch and the core-shell elasticity. We also determine the shape phase diagram for a spherical void bounded by a fixed-area membrane.
\end{abstract}

\maketitle

\section{Introduction}
There has been longstanding interest
in the mechanical instabilities of
core-shell systems,
comprising an elastic sphere on the inside,
surrounded by and attached to an elastic exterior shell.
Although idealized, such a model has
been proposed to underlie
myriad buckled or wrinkled biological morphologies,
such as those of
fruits and vegetables \cite{doi:10.1073/pnas.0810443105,Yin2009,PhysRevLett.106.234301},
insect eggs \cite{Munguira2015},
pollen grains \cite{Katifori2010,Radja2019},
neutrophils and B cells  \cite{TingBeall1993, ROWAT20138610,C1SM06637D}, 
mammalian brains
\cite{Tallinen2014,PhysRevE.92.032701},
and growing tumors \cite{PhysRevLett.110.158102}.
In addition to these biological examples,
swelling gels often show similar mechanical instabilities \cite{Tanaka1987,C2SM25617G,PhysRevLett.107.018103,PhysRevApplied.6.064010}, as do inorganic core-shell systems
\cite{Li2005,PhysRevLett.100.036102}.

%

%
%

To fully discuss spherical core-shell
systems,
it is important to properly understand the elasticity
of the spherical core.
For an isotropic material with Poisson ratio,
$\nu$,
in mechanical equilibrium,
according to linear elasticity
theory,
the elastic displacement field, $\mathbf{u}$, must satisfy
    \begin{equation} \label{eom}
    \nabla (\nabla \cdot \mathbf{u})+(1-2\nu)\nabla^2 \mathbf{u}=0,
    \end{equation}
which is the statement that the force density is zero everywhere within the material of the spherical core.
Eq.~\ref{eom} plays an analogous role in
elasticity theory to that  played
 in electrostatics by Laplace's equation,
whose solutions are well-known to be the regular
and irregular solid harmonics, namely $r^l Y_l^m(\theta,\phi)$
and $r^{-l-1} Y_l^m(\theta,\phi)$, respectively.
From this point of view, it is surprising that
analytic solutions of Eq.~\ref{eom} in near spherical
situations have been little discussed.
The corresponding elastic energies of these solutions
also remain unknown, as far as we are aware.
Ref.~\onlinecite{PhysRevE.88.052404} sought to
remedy this situation,
by, first,
solving Eq.~\ref{eom} for an elastic sphere subject to the
boundary condition that the sphere's surface is
displaced radially with an amplitude
given by a real spherical harmonic, and, then, by calculating
the corresponding elastic energies.
However, as  described below, we disagree with
Ref.~\onlinecite{PhysRevE.88.052404}'s
result that the elastic energy
depends on the spherical harmonic index, $m$.

The goal of this paper is threefold:
(1)  to find the displacement
field both within a sphere, with a real-spherical-harmonic surface
displacement, and outside a spherical void, with a real-spherical-harmonic surface
displacement; (2) to calculate corresponding bulk elastic energies;
and (3) to use the resultant elastic energy to determine the shape phase diagram both of a core-shell system, comprising an elastic sphere, attached within a
membrane of fixed area \cite{PhysRevE.88.052404},
and of a spherical void, which is lined by
a membrane of fixed area, that is attached to
the surrounding elastic medium.
A number of recent contributions have focused on post-buckling pattern selection in core-shell
systems,
which depends on non-linear
effects \cite{PhysRevLett.100.036102,PhysRevLett.106.234301,C1SM06637D,Breid2013,Tallinen2014,Stoop2015,Radja2019,Xu2020,Xu2022}.
However, such phenomena lie
beyond our scope, which is confined to linear elasticity only.

The outline of the paper is as follows.
By exploiting well-known properties of the solid harmonics, 
we first present a straightforward,
direct method for
solving Eq.~\ref{eom} in general, near-spherical situations,
 both for spheres (Sec.~\ref{reg.sol}) and spherical voids (Sec.~\ref{irreg.sol}).
Then, we fit the general solutions
to boundary conditions corresponding to
a spherical core (Sec.~\ref{bcs}) or a spherical void (Sec.~\ref{bcs2}), whose surface is displaced radially
with an amplitude given by a real spherical harmonic.
In Sec.~\ref{ebulk}, we calculate the bulk elastic energies
corresponding to these boundary conditions.
We provide analytic expressions for 
the energies for any value of the spherical harmonic degree,
$l$, Poisson ratio, $\nu$, and shear modulus, $\mu$.
The elastic energies are independent of the spherical
harmonic index, $m$.
In Sec. \ref{coreshell}, following Ref.~\onlinecite{PhysRevE.88.052404},
we revisit the buckling instability
experienced by a core-shell system comprising an elastic sphere,
attached within a membrane of fixed area,
that occurs when the area of the membrane sufficiently exceeds the area of the unstrained sphere.
We determine the phase diagram of the core-shell sphere's shape,
specifying what value of $l$ is realized as a function of area mismatch and sphere and membrane elasticity.
Similarly, we also determine the analogous shape phase
diagram for a spherical void bounded by a fixed-area membrane.
A Mathematica notebook containing all of our calculations is
available at Github \cite{URL}.

\section{Regular solution for spheres} \label{reg.sol}
To find solutions to Eq.~\ref{eom},
applicable to (slightly deformed) spheres,
we first introduce two trial
functions, that when summed together with appropriate relative weighting,
indeed satisfy Eq.~\ref{eom}.
To this solution, we  then add an
additional trial function that satisfies Eq.~\ref{eom} on its own,
yielding a final result, that can be conveniently matched to
the applicable boundary conditions.

Trial function 1 takes the form
\begin{equation}
  \mathbf{u_1}=a r^2\nabla(r^l Y_l^m), 
  \label{EQ2}
\end{equation}
where $a$ is a constant.
Eq.~\ref{EQ2} converges at $r=0$, and eventually will be part of the so-called
regular solution.
It follows from Eq.~\ref{EQ2} that,
    \begin{equation} \label{EQ3}
    \begin{split}
    \nabla \cdot \mathbf{u_1}&=a(\nabla r^2)\cdot \nabla(r^l Y_l^m)+ar^2\nabla^2(r^l Y_l^m) \\
    &=2lar^lY_l^m,
    \end{split}
    \end{equation}
and, in turn, that
    \begin{equation} \label{EQ4}
    \nabla(\nabla \cdot \mathbf{u_1})=2la\nabla(r^lY_l^m).
    \end{equation}
We also have that
    \begin{equation} \label{EQ5}
    \nabla^2\mathbf{u_1}=2(2l+1)a\nabla(r^lY_l^m).
    \end{equation}
    Combining Eq.~\ref{EQ4} and Eq.~\ref{EQ5} yields
    \begin{equation} \label{1result}
    \nabla (\nabla \cdot \mathbf{u_1})+(1-2\nu)\nabla^2 \mathbf{u_1}=(2l+2(1-2\nu)(2l+1))a\nabla(r^lY_l^m)
    \end{equation}
    Thus, Eq.~\ref{eom} produces a non-zero result for $\mathbf{u}_1$, and another trial function is needed to cancel $\mathbf{u}_1$ in order to satisfy Eq.~\ref{eom}. \par
To this end, we introduce trial function 2:
\begin{equation}
    \label{EQ7}
\mathbf{u_2}=\mathbf{b} r^{l+1} Y_{l+1}^m,
\end{equation}
where $\mathbf{b}=(b_x,b_y,b_z)$ is a constant vector.
 Then,
\begin{equation}
\begin{split}
    \nabla \cdot \mathbf{u_2}&=\mathbf{b}\cdot \nabla(r^{l+1} Y_{l+1}^m) \\
    &=\alpha r^lY_l^{m+1}+\beta r^lY_l^{m}+\gamma r^lY_l^{m-1}
    \end{split}
\end{equation}
    where $\alpha$, $\beta$ and $\gamma$ are all known quantities,
    given explicitly in the Appendix
    (Eq.~\ref{dYdx}, Eq.~\ref{dYdy}, and Eq.~\ref{dYdz}, respectively).
    Since
     $\nabla^2\mathbf{u_2}=0$,
   we have that
   \begin{widetext}
    \begin{equation}
    \label{EQ9}
     \nabla (\nabla \cdot \mathbf{u}_2)+(1-2\nu)\nabla^2 \mathbf{u}_2=
        \nabla(\nabla \cdot \mathbf{u_2})=\alpha \nabla(r^lY_l^{m+1})+\beta \nabla(r^lY_l^{m})+\gamma \nabla(r^lY_l^{m-1}).
    \end{equation}
    \end{widetext}
The terms on the right-hand side of Eq.~\ref{EQ9} are of the same form as the right-hand side of Eq. \ref{1result}, except for the appearance of additional terms with spherical harmonic indeces equal to $m \pm 1$.
However, we can use a modified version of $\mathbf{u_1}$,
augmented to cancel all three terms arising from $\mathbf{u}_2$. Because
$\nabla (r^l Y_l^m)$
satisfies Eq.~\ref{eom} on its own, we can also add additional terms of this form to $\mathbf{u_1}$, with a view to
the solution for a surface displacement given by a single
spherical harmonic.
Specifically, we can pick
\begin{equation}
\begin{split}   \mathbf{u_1'}=&a_1(r^2-R^2)\nabla(r^{l}Y_l^{m+1})\\
    &+a_0(r^2-R^2)\nabla(r^{l}Y_l^{m})\\
    &+a_{-1}(r^2-R^2)\nabla(r^{l}Y_l^{m-1}),
\end{split}
\end{equation}
where $R$ is the radius of the undeformed sphere,
\begin{equation}
     a_1 =\frac{-\alpha}{2l+2(1-2\nu)(2l+1)},
     \end{equation}
     \begin{equation}
    a_0 =\frac{-\beta}{2l+2(1-2\nu)(2l+1)},
     \end{equation}
     and
     \begin{equation}
    a_{-1} =\frac{-\gamma}{2l+2(1-2\nu)(2l+1)}.
\end{equation}
By construction, Eq.~\ref{eom} is now satisfied by
\begin{equation}
\begin{split}
    \mathbf{u}_{lm}=&\mathbf{u}_1'+\mathbf{u}_2\\
    =&a_1(r^2-R^2)\nabla(r^{l}Y_l^{m+1})\\
    &+a_0(r^2-R^2)\nabla(r^{l}Y_l^{m})\\
    &+a_{-1}(r^2-R^2)\nabla(r^{l}Y_l^{m-1})\\
    &+(b_x,b_y,b_z)r^{l+1} Y_{l+1}^m.
    \label{EQ20}
    \end{split}
\end{equation}
Using the expressions for $\alpha$, $\beta$, and $\gamma$, given in the Appendix, we have
\begin{equation}
    a_{1}=\frac{(b_x-i b_y) \sqrt{\frac{(2 l+3) (l-m+1)!}{(l+m+1)!}}}{2(l (8 \nu -6)+4\nu -2) \sqrt{\frac{(2 l+1) (l-m-1)!}{(l+m+1)!}}},
\end{equation}
\begin{equation}
    a_{0}=\frac{b_z (l+m+1) \sqrt{\frac{(2 l+3) (l-m+1)!}{(l+m+1)!}}}{(l (8 \nu -6)+4 \nu -2)\sqrt{\frac{(2 l+1) (l-m)!}{(l+m)!}}},
\end{equation}
and
\begin{equation}
    a_{-1}=-\frac{(b_x+i b_y) (l+m) (l+m+1) \sqrt{\frac{(2 l+3) (l-m+1)!}{(l+m+1)!}}}{2(l
(8 \nu -6)+4 \nu -2) \sqrt{\frac{(2 l+1) (l-m+1)!}{(l+m-1)!}}}.
\end{equation}

While $\mathbf{u}_1'$ involves
spherical harmonics of degree $l-1$, by contrast, $\mathbf{u}_2$ involves spherical
harmonics of degree $l+1$. Thus, solutions to Eq.~\ref{eom} necessarily involve at least one pair of
values of $l$ that differ by 2.
We also see that solutions to Eq.~\ref{eom} naturally involve three consecutive
values of $m$.

\par
For $r=R$, we see that only $\mathbf{u_2}$ survives.
Because  $\mathbf{u_2}$ involves a single spherical harmonic,
this approach facilitates matching surface displacements,
that are given by a spherical harmonic or a sum of spherical harmonics.

\section{Irregular solution for spherical voids}
\label{irreg.sol}
Using an analogous procedure to that followed in Sec.~\ref{reg.sol},
we can also find a solution that remains finite as $r \rightarrow \infty$,
namely the irregular solution, which
is applicable within elastic material surrounding a spherical void.
To this end, we again introduce two trial functions,
\begin{equation}
\begin{split}
    \mathbf{v_1}=&a_1(r^2-R^2)\nabla(r^{-l-1}Y_l^{m+1})\\
    &+a_0(r^2-R^2)\nabla(r^{-l-1}Y_l^{m})\\
    &+a_{-1}(r^2-R^2)\nabla(r^{-l-1}Y_l^{m-1})\\,
    \end{split}
\end{equation}
and 
\begin{equation}
\begin{split}
    \mathbf{v_2}&=\mathbf{b} r^{-l}Y_{l-1}^m \\
    &=(b_x,b_y,b_z)r^{-l}Y_{l-1}^m.
    \end{split}
\end{equation}
To ensure that $\mathbf{v}_1+\mathbf{v}_2$ is a solution to Eq.~\ref{eom},
we must pick
\begin{eqnarray}
a_ 0 &= \frac {b_z\sqrt {\frac {-1 + 2 l} {3 + 2 l}}\sqrt {l^2 - 
      m^2}} {2\sqrt {\frac {1 + 2 l} {3 + 2 l}} (-2 - 3 l + 2 \nu + 
      4 l \nu)}, \\
a_ 1 &= -\frac {(b_x - 
       ib_y)\sqrt {\frac {-1 + 2 l} {3 + 2 l}}\sqrt {(l + m) (1 + l + 
        m)}} {4\sqrt {\frac {1 + 2 l} {3 + 2 l}} (-2 - 3 l + 2 \nu + 
      4 l \nu)}, \\
a_ {-1} &= \frac {(b_x + 
       ib_y)\sqrt {\frac {-1 + 2 l} {3 + 2 l}}\sqrt {l + l^2 - 
      2 l m + (-1 + m) m}} {4\sqrt {\frac {1 + 2 l} {3 + 2 l}} (-2 - 
      3 l + 2 \nu + 4 l \nu)},
\end{eqnarray}
so that the contributions of the two trial functions to Eq.~\ref{eom} cancel.

The irregular solution is
\begin{equation}
\begin{split}
    \mathbf{v}_{lm}=&\mathbf{v_1}+\mathbf{v_2}\\
    =&a_1(r^2-R^2)\nabla(r^{-l-1}Y_l^{m+1})\\
    &+a_0(r^2-R^2)\nabla(r^{-l-1}Y_l^{m})\\
    &+a_{-1}(r^2-R^2)\nabla(r^{-l-1}Y_l^{m-1})\\
    &+(b_x,b_y,b_z)r^{-l}Y_{l-1}^m.
    \label{EQ27}
    \end{split}
\end{equation}
Two values of $l$ are involved
in the irregular solution too,
and only the coefficients in $\mathbf{v_2}$
need be considered to fit boundary conditions at $r=R$.
\begin{figure*}[h]
    \centering
    \includegraphics[width=.49\textwidth]{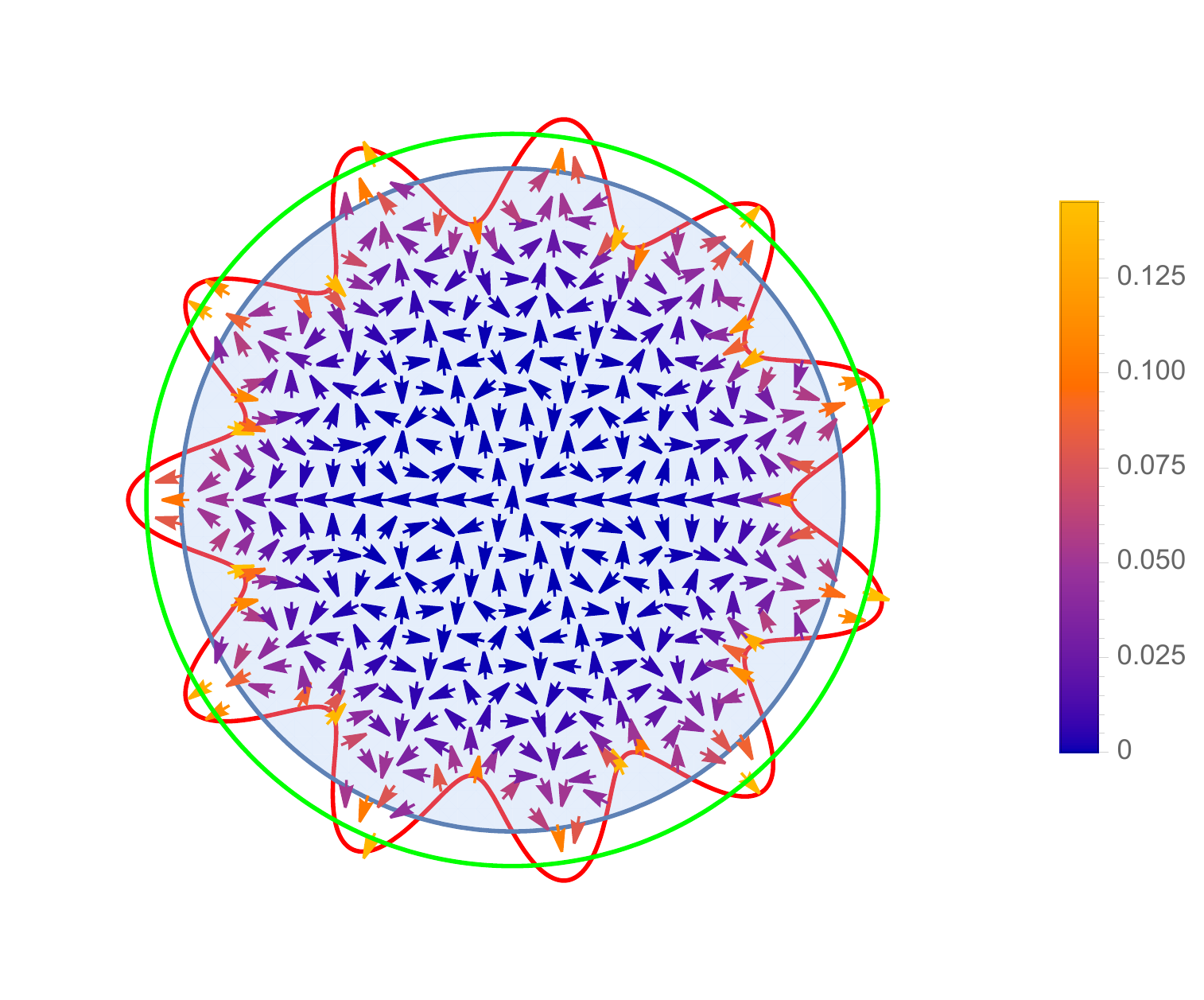}
    \includegraphics[width=.49\textwidth]{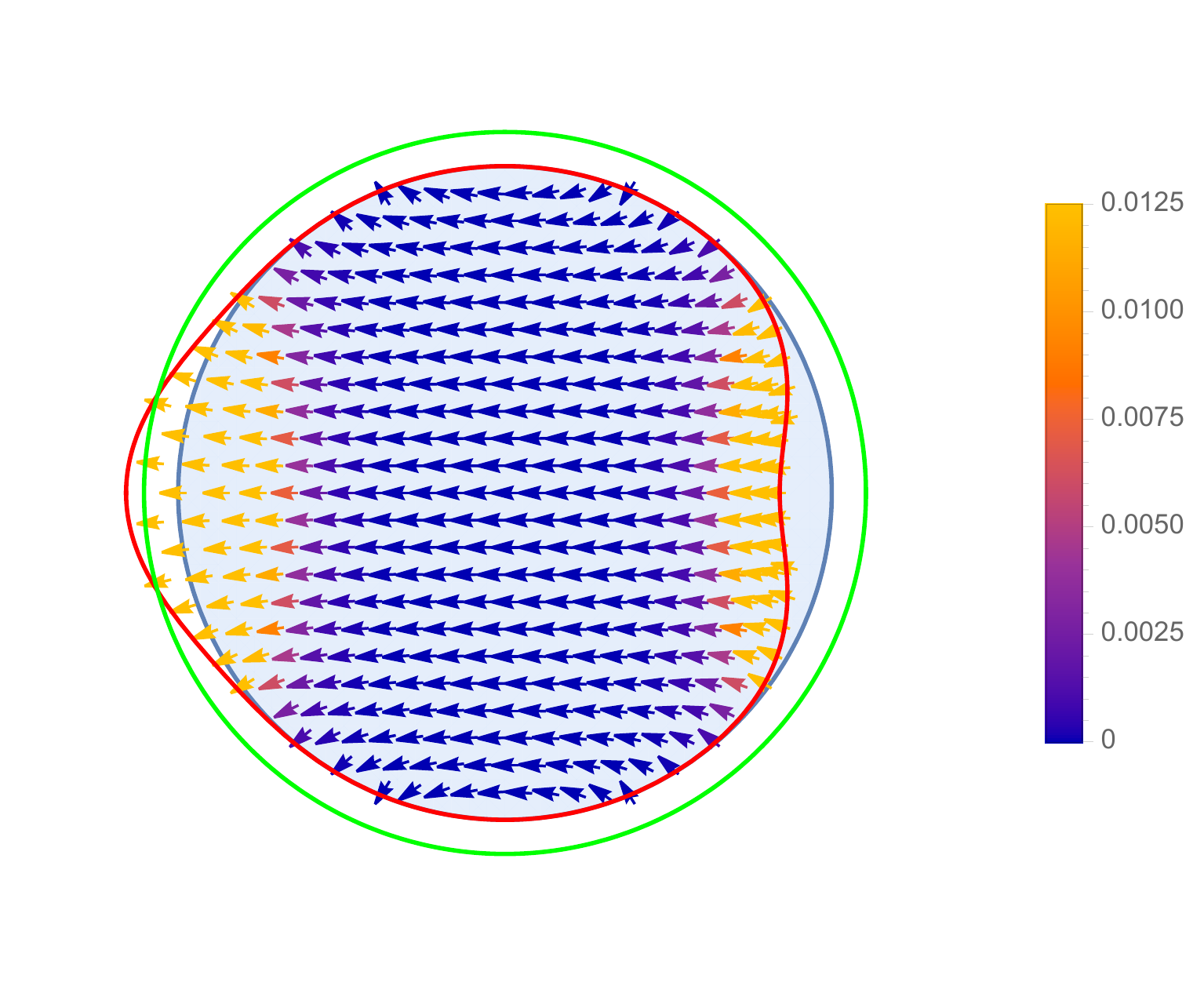}
\caption{Plot of the shape and displacement field within the $xy$-plane (left) and the $xz$-plane (right) of a sphere with a surface deformation given by a spherical harmonic with degree $l=11$ and index $m=11$ and $\nu=0.3$.
    The direction of the arrows represents the direction of the elastic displacements within the sphere. The color of the arrows represents the magnitude of these displacements. The buckled shape, represented by the red curve, has a spherical harmonic amplitude of $g=0.20$,
 corresponding
    to the 
    excess area at the transition from the isotropically-expanded phase to the buckled phase (Sec. \ref{coreshell}). The smaller blue circle represents the undeformed sphere. The larger green circle represents the isotropically-expanded sphere with the same surface area as the buckled shape. 
    }
    \label{spherexy}
\end{figure*}

\begin{figure*}[h]
    \centering
    \includegraphics[width=.49\textwidth]{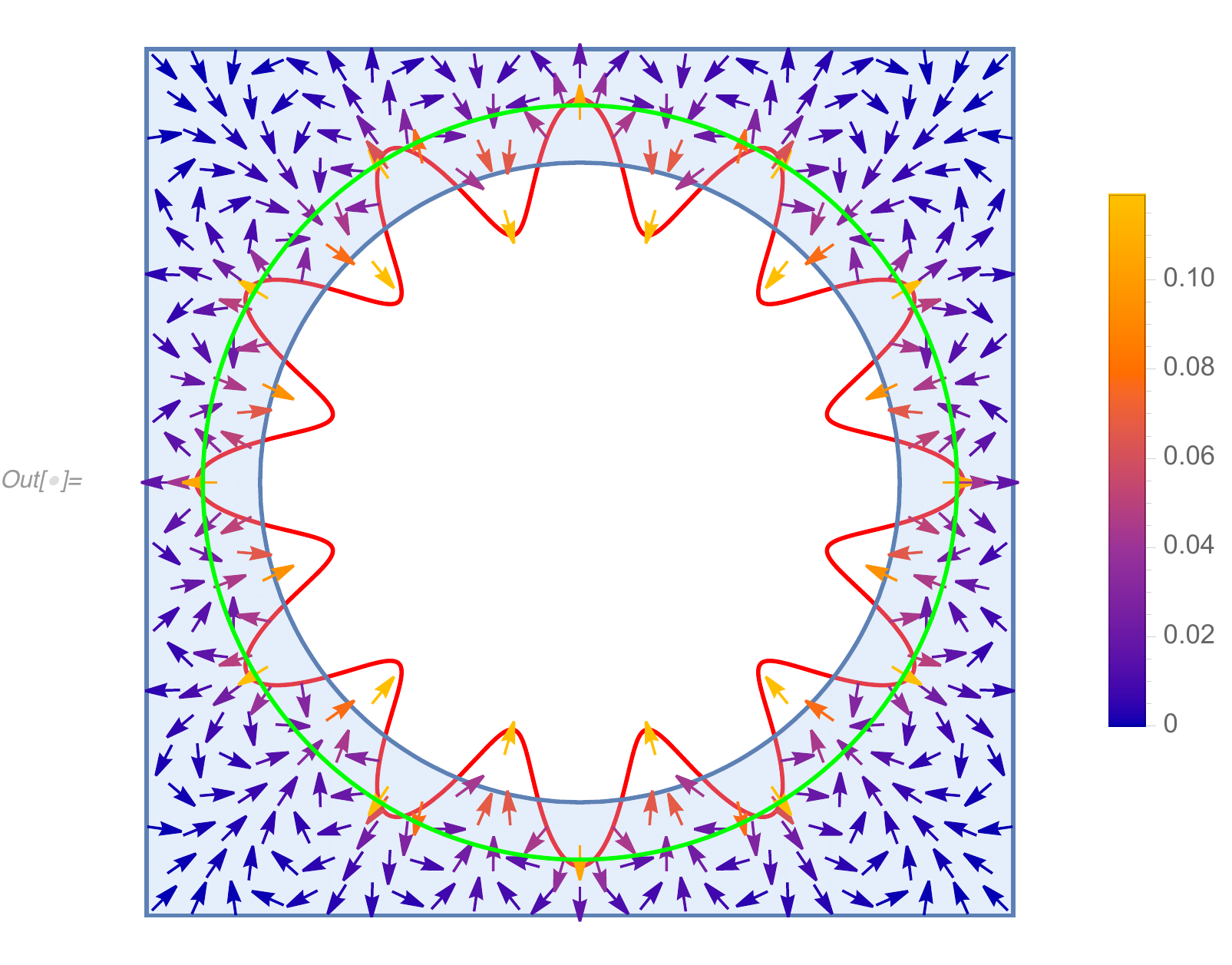}
     \includegraphics[width=.49\textwidth]{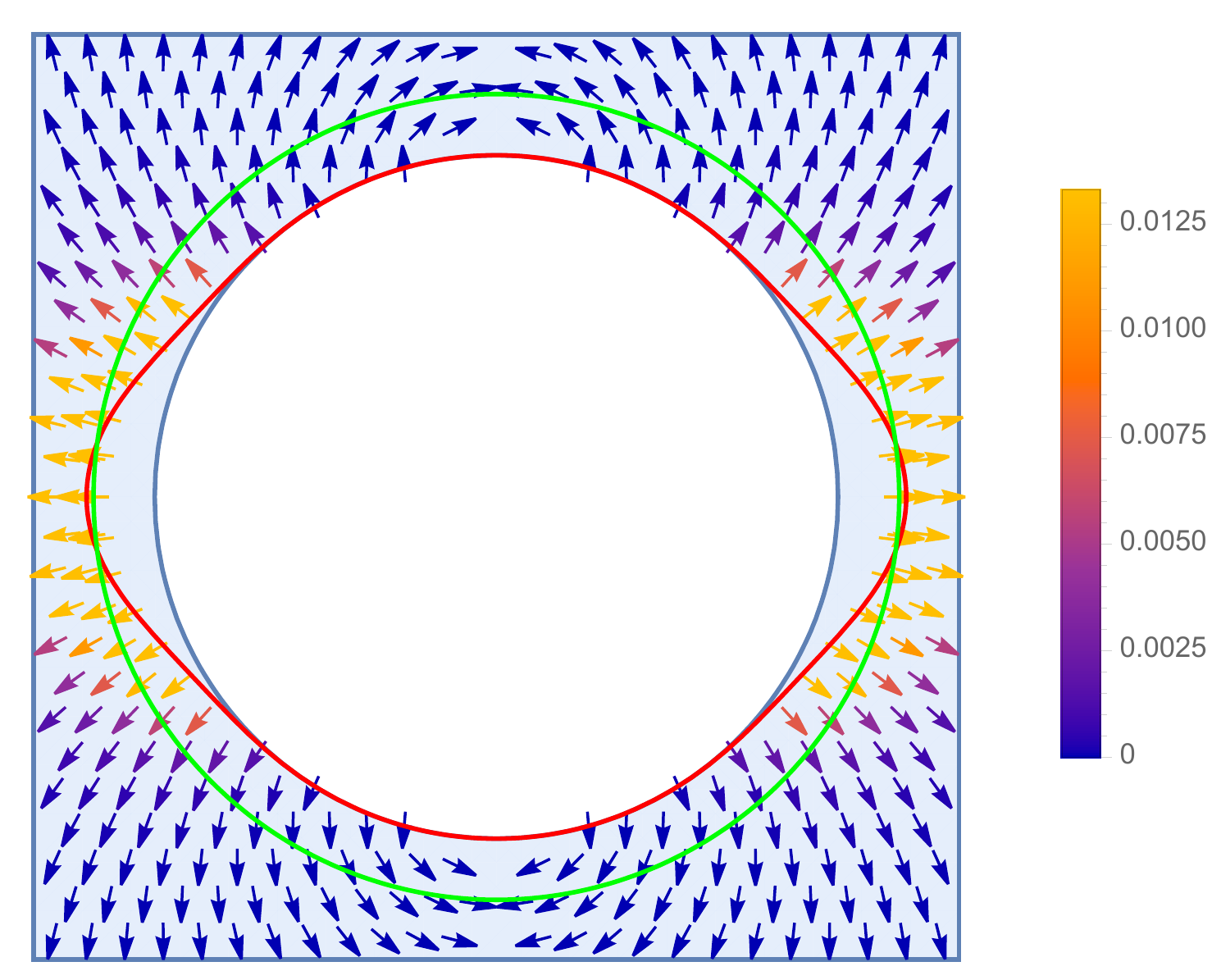}
    \caption{Plot of the shape and displacement field within the $xy$-plane (left) and the $xz$-plane (right) of a spherical void with a surface deformation given by a spherical harmonic with degree $l=12$ and index $m=12$ and $\nu=0.3$.
    The direction of the arrows represents the direction of the elastic displacements within the sphere. The color of the arrows represents the magnitude of these displacements. The buckled shape, represented by the red curve, has a spherical harmonic amplitude of $g=0.25$,
 corresponding
    to the 
    excess area at the transition from the isotropically-expanded phase to the buckled phase (Sec. \ref{coreshell}). The smaller blue circle represents the undeformed sphere. The larger green circle represents the isotropically-expanded sphere with the same surface area as the buckled shape. 
   }
    \label{voidxy}
\end{figure*}

\section{Sphere with a spherical harmonic shape deformation}
\label{bcs}
Next, we consider a (slightly deformed) sphere,
whose shape deviates from a perfect sphere by a single, real spherical harmonic, $Y_{lm}$, defined as 
\begin{equation}
    Y_{lm}=\frac{1}{\sqrt{2}}[Y_l^m+(-1)^mY_l^{-m}]
\end{equation}
for $m>0$ and as $Y_{l0}=Y_l^{0}$ for $m=0$.
The amplitude of the  displacement
of the elastic medium immediately
behind the surface is proportional to the surface displacement.
We furthermore suppose that this displacement is
directed along the radial direction.
Thus, the relevant boundary condition is that
the displacement at the surface is
 \begin{equation}
 \mathbf{u}(R) = g R Y_{lm} \hat{\mathbf{r}},
 \label{VectorBCs}
 \end{equation}
 where $g$ is a dimensionless measure of the amplitude of the surface displacement.
 The radial unit vector,
$\hat{\mathbf{r}}=(\sin{\theta}\cos{\phi},\sin{\theta}\sin{\phi},\cos{\theta})$, may be
 expressed in terms of
 $Y_1^1$, $Y_{1}^0$ and $Y_{1}^{-1}$:
 \begin{widetext}
\begin{equation}
\begin{split}
\hat{\mathbf{r}}=(&-\sqrt{\frac{2 \pi }{3}} \left(Y_1^{1}(\theta , \phi ) -Y_1^{-1}(\theta ,\phi )\right),
i \sqrt{\frac{2 \pi }{3}}
\left(Y_1^{-1}(\theta ,\phi )+Y_1^{1}(\theta ,\phi )\right),
2\sqrt{\frac{\pi }{3}} Y_1^{0}(\theta ,\phi )),
\end{split}
\end{equation}
implying that Eq.~\ref{VectorBCs} consists of
pairwise products of spherical harmonics.
It is well-known,
however, that  pairwise products of spherical harmonics may be expressed as a linear combination of spherical harmonics
with degrees and indeces and weights, specified by the 
 Wigner 3-j symbols. Thus,
we find that 
 Eq.~\ref{VectorBCs}
 contains spherical harmonics
 with degree $l \pm 1$,
 and indeces $m \pm 1$ for the
 $x$- and $y$-components,
 and index $m$ for
 the $z$-component,
 and the complex conjugates of these
 terms,
 which is a total of twelve
 spherical harmonics each with a
 different combination of $l$
 and $m$ than the others (Table~\ref{Table1}).
 This form of  Eq.~\ref{VectorBCs}
 is given in the
 accompanying Mathematica notebook.
To satisfy these boundary conditions,
we must select a solution
that
is
a superposition of twelve
${\bf u}_{lm}$'s
containing the values of $l$
and $m$ needed, and we must
set the
components of $\bf b$ for
each ${\bf u}_{lm}$
in the superposition
equal to the coefficient
of the corresponding
spherical harmonic in Eq.~\ref{VectorBCs}.
Thus, we find the following solution
for spheres:
\begin{equation}
    u_x=R\sum_{m'=m\pm1} (A_{m'}(r^2-R^2)r^{l-1}Y_{l-1}^{m'}+B_{m'}r^{l+1} Y_{l+1}^{m'}+C_{m'}r^{l-1} Y_{l-1}^{m'})+\text{c.c.}
        \label{EQ32}
\end{equation}
\begin{equation}
    u_y=R\sum_{m'=m\pm1} (D_{m'}(r^2-R^2)r^{l-1}Y_{l-1}^{m'}+E_{m'}r^{l+1} Y_{l+1}^{m'}+F_{m'}r^{l-1} Y_{l-1}^{m'})+\text{c.c.}
        \label{EQ33}
\end{equation}
\begin{equation}
    u_z=R\left(G(r^2-R^2)r^{l-1}Y_{l-1}^{m}+Hr^{l+1} Y_{l+1}^{m}+Ir^{l-1} Y_{l-1}^{m}\right)+\text{c.c.}
    \label{EQ34}
\end{equation}
\end{widetext}
where the coefficients
($A_{m'}$, $B_{m'}$, {\em etc.}) are all known functions of $l,\ m,\ R,\ g$
and $\nu$, and are given in Appendix
\ref{AppendixC}.
It turns out that the coefficients vanish
for all terms of the form $(r^2-R^2) Y_{l-3}^{m'}$, that would otherwise appear in Eqs.~\ref{EQ32}, \ref{EQ33}, and \ref{EQ34}. \par
The displacement field ($\mathbf{u}$) for a
sphere with a
spherical harmonic surface
deformation with $l=11$ and $m=11$ is illustrated in Fig.~\ref{spherexy} for $g=0.22$ and $\nu=0.3$.
This representation
shows how the interior of the original sphere (blue, smaller circle) is deformed to the buckled shape (red curve). The larger green circle
has the same surface area as the buckled shape, 
and is included for reference.

\section{Spherical voids with a spherical harmonic shape deformation}
\label{bcs2}
Similarly to Sec.~\ref{bcs},
our solution for spherical voids is:
\begin{widetext}
\begin{equation}
    v_x=R\sum_{m'=m\pm1} (J_{m'}(r^2-R^2)r^{-2-l}Y_{l+1}^{m'}+K_{m'}r^{-l} Y_{l-1}^{m'}+L_{m'}r^{-2-l} Y_{l+1}^{m'})+\text{c.c.}    
\end{equation}

\begin{equation}
    v_y=R\sum_{m'=m\pm1} (M_{m'}(r^2-R^2)r^{-2-l}Y_{l+1}^{m'}+N_{m'}r^{-l} Y_{l-1}^{m'}+O_{m'}r^{-2-l} Y_{l+1}^{m'})+\text{c.c.}    
\end{equation}

\begin{equation}
    v_z=R\left( P(r^2-R^2)r^{-2-l}Y_{l+1}^{m}+Qr^{-l} Y_{l-1}^{m}+Sr^{-2-l} Y_{l+1}^{m}\right)+\text{c.c.}    
\end{equation}
\end{widetext}
where the coefficients here
are given in Appendix
\ref{AppendixD}.
The displacement field for a
spherical void ($\mathbf{v}$) with a
spherical harmonic surface
deformation with $l=12$ and $m=12$
 is similarly plotted in Fig.~\ref{voidxy}
 for 
$\nu=0.3$ and $g=0.25$.

  %


\section{Bulk elastic energies} \label{ebulk}
Elasticity theory informs us that the 
elastic energy density, $w$, can be directly calculated from the derivatives of the displacement $\mathbf{u}$, namely the strains,
$\epsilon_{ij}=\frac{1}{2}(\partial_{i}u_j+\partial_{j}u_i)$:
    \begin{equation} 
    \begin{split}
    w=&\mu[\frac{1-\nu}{1-2\nu}(\epsilon_{xx}^2+\epsilon_{yy}^2+\epsilon_{zz}^2)\\
    &+\frac{2\nu}{1-2\nu}(\epsilon_{xx}\epsilon_{yy}+\epsilon_{yy}\epsilon_{zz}+\epsilon_{zz}\epsilon_{xx})\\
    &+2(\epsilon_{xy}^2+\epsilon_{yz}^2+\epsilon_{zx}^2)].
    \label{energy_density}
    \end{split}
    \end{equation}
Then,  to find the total bulk energy, $W$,
we must integrate the energy density over the volume of the sphere (or over the volume outside the spherical void).

Using Eqs.~\ref{EQ32}, \ref{EQ33}, and \ref{EQ34}, in conjunction
with Eqs. \ref{dYdx}, \ref{dYdy}, and \ref{dYdz} from
Appendix \ref{appendixA}, we can calculate each strain component
with the result that each strain component comprises
a sum of up to twenty spherical harmonics:
\begin{equation} \label{epsilon}
    \epsilon_{ij}=\sum_{\substack{l'=l-2,l \\ m'=\pm{m},\pm{m}\pm{1},\pm{m}\pm{2}}} d_{l',m'} Y_{l'}^{m'}
\end{equation}
where $d_{l',m'}$ are the coefficients of $Y_{l'}^{m'}$ in $\epsilon_{ij}$ and depend on cartesian coordinates $i$ and $j$
(Table~\ref{Table1}).

The spherical harmonics are orthogonal and normalized,
that is,
 \begin{equation}
    \int Y_l^m (Y_{l'}^{m'})^* d\Omega=\delta_{ll'} \delta_{mm'}
\end{equation}
where $(Y_l^{m})^*$ is the complex conjugate of $Y_l^{m}$.
Since $(Y_l^{m})^* = (-1)^mY_l^{-m}$, it follows that
\begin{equation}
    \int Y_l^m Y_{l'}^{-m'} d\Omega=(-1)^m \delta_{ll'}\delta_{mm'}.
    \label{EQ41}
\end{equation}
We can use this result to facilitate integration of the energy density over angles by first representing
 $\epsilon_{ij}$ as two vectors, each of 10 components,
 one corresponding to spherical harmonics of degree $l$ and the other corresponding
 to spherical harmonics of degree $l-2$ ($l$ and $l+2$ for irregular solution):
\begin{widetext}
\begin{equation}\label{l}
    \mathbf{d}_{\epsilon_{ij}}=(d_{l',m+2},d_{l',m+1},d_{l',m},d_{l',m-1},d_{l',m-2},d_{l',-m+2},d_{l',-m+1},d_{l',-m},d_{l',-m-1},d_{l',-m-2})
\end{equation}
where $l'=l-2,\ l$.
Next, for each of $l$ and $l-2$,  we construct a $10 \times 10$ matrix, whose
entries derive from the left hand side of Eq.~\ref{EQ41}:
\begin{equation}
M =
\left(
\begin{array}{cccccccccc}
 \delta _{-2,m} & 0 & -\delta _{-1,m} & 0 & \delta _{0,m} & 0 & 0 & 0 & 0 & (-1)^m \\
 0 & \delta _{-1,m} & 0 & -\delta _{0,m} & 0 & 0 & 0 & 0 & (-1)^{m+1} & 0 \\
 -\delta _{-1,m} & 0 & \delta _{0,m} & 0 & -\delta _{1,m} & 0 & 0 & (-1)^m & 0 & 0 \\
 0 & -\delta _{0,m} & 0 & \delta _{1,m} & 0 & 0 & (-1)^{m+1} & 0 & 0 & 0 \\
 \delta _{0,m} & 0 & -\delta _{1,m} & 0 & \delta _{2,m} & (-1)^m & 0 & 0 & 0 & 0 \\
 0 & 0 & 0 & 0 & (-1)^m & \delta _{2,m} & 0 & -\delta _{1,m} & 0 & \delta _{0,m} \\
 0 & 0 & 0 & (-1)^{m+1} & 0 & 0 & \delta _{1,m} & 0 & -\delta _{0,m} & 0 \\
 0 & 0 & (-1)^m & 0 & 0 & -\delta _{1,m} & 0 & \delta _{0,m} & 0 & -\delta _{-1,m} \\
 0 & (-1)^{m+1} & 0 & 0 & 0 & 0 & -\delta _{0,m} & 0 & \delta _{-1,m} & 0 \\
 (-1)^m & 0 & 0 & 0 & 0 & \delta _{0,m} & 0 & -\delta _{-1,m} & 0 & \delta _{-2,m} \\
\end{array}
\right)
.
\label{matrix}
\end{equation}
\end{widetext}
It then follows that the required integrals over angles now
correspond to matrix multiplication: 
\begin{equation}
   \int \epsilon_{ij} \epsilon_{pq} d \Omega= \mathbf{d}_{\epsilon_{ij}} M {\mathbf{d}_{\epsilon_{pq}}}^T,
\end{equation}
where ${\mathbf{d}_{\epsilon_{pq}}}^T$ is the transpose
of ${\mathbf{d}_{\epsilon_{pq}}}.$
The integrals over $r$  must be done explicitly:
\begin{equation}
    \int_0^R  \mathbf{d}_{\epsilon_{ij}} M {\mathbf{d}_{\epsilon_{pq}}}^T r^2 dr
\end{equation}
for spheres and
\begin{equation}
    \int_R^\infty  \mathbf{d}_{\epsilon_{ij}} M {\mathbf{d}_{\epsilon_{pq}}}^T r^2 dr
\end{equation}
for spherical voids.
Even though several of the matrix elements of matrix, M (Eq.~\ref{matrix}),
explicitly depend on particular values of the spherical harmonic index,
$m$, 
both the total elastic energy and the elastic
energy within a shell at radius $r$
are independent of $m$,  that is, all shapes with the same $l$ have the same energy. 
Because the expression for the elastic energy
is invariant under rotations,
we can understand the $m$-independence of the elastic energy by realizing that with the given boundary conditions --
a radial surface displacement with an amplitude proportional to
a spherical harmonic -- the elastic energy
is a functional of the spherical-harmonic surface shape.
Since each spherical harmonic is an irreducible representation
of the rotation group, the elastic energy must therefore
be independent of $m$.

 For general values of $l$, the expression for the bulk elastic energy appears unwieldly
(as can be seen from the Mathematica notebook).
However, for any specific value of $l$,
the elastic energy reduces to a remarkably simple form.
Examination of this energy  for values of $l$ from 1 to 25,
using Mathematica's {\tt FindSequenceFunction},
indicates that the bulk elastic energies
are given
for general $l$ by
\begin{equation}
    W=g^2 \mu R^3 \left( \frac{\left(2 l^2-3l-1\right)\nu -\left(2l^2-l+1\right)}{2
(2 l+1)\nu -(3 l+1)}
\right)\\
\label{EQ46}
\end{equation}
for spheres,
and
\begin{equation}
W=g^2 \mu R^3 \left( \frac{\left(4+7 l+2 l^2\right)\nu -\left(4+5
l+2 l^2\right)}{2 (1+2 l)\nu -(2+3 l)}
\right)
\label{EQ47}
\end{equation}
 for spherical voids.
Eq.~\ref{EQ46} and Eq.~\ref{EQ47}
are key results of this paper.

Fig. \ref{energy_density_r}
and Fig. \ref{energy_density_i}
present the energy density,
averaged over angles,
within a shell at radius $r$  for spheres and spherical voids, respectively.
Inspection of Fig. \ref{energy_density_r} and Fig. \ref{energy_density_i}
makes it clear that for
increasing
$l$,  most of the energy density, displacement and strain
is confined to an increasingly narrow near-surface layer.
 In Fig. \ref{energy_density_r}
 for spheres,
 each curve displays a peak at a radius less than $R$,
 which appears progressively closer to the surface for 
 progressively larger $l$ values.
 By contrast, in Fig. \ref{energy_density_i}, the curves
 for spherical voids
 appear to decrease monotonically as $r$ increases.

For boundary conditions, described by the sum of two spherical
harmonics, $Y_{lm}$ and $Y_{l'm'}$, the solution for the displacement $\mathbf{u}$ is the sum of the two solutions, satisfying boundary conditions described by $Y_{lm}$ and $Y_{l'm'}$ separately.
This result is inevitable given that Eq.~\ref{eom} is linear in $\bf u$.
We furthermore find that the corresponding bulk elastic energy is also additive, {\em i.e.} the energy for the two-spherical-harmonic boundary condition,
$Y_{lm}+Y_{l'm'}$,
is the sum of the energy for boundary condition, $Y_{lm}$, and the energy for boundary condition, $Y_{l'm'}$.
The reason is clear for cases in which  $l$ and $l'$ are far apart, because from
Eq. \ref{epsilon},  ${\bf u}_{lm}$ and ${\bf u}_{l'm'}$ then
have no spherical harmonics in common.
However, even in cases where $l-l'=2$, so that  the same spherical harmonics may
appear in both ${\bf u}_{lm}$ and ${\bf u}_{l'm'}$, we find that the energy is additive.

Finally, as the alternatives to
a buckled sphere and a buckled spherical void, we consider the elastic energy of a isotropically expanded sphere
and an isotropically expanded spherical void.
In the case of an isotropically expanded sphere,
the displacement is $\mathbf{u}=
g{\bf r}$ (which
also satisfies Eq.~\ref{eom}),
so that $u_x=gx, u_y=gy, u_z=gz$, and $\epsilon_{xx}=\epsilon_{yy}=\epsilon_{zz}=g$ while $\epsilon_{ij}=0$
for $i \neq j$.
Substituting these results for the strains into
Eq.~\ref{energy_density}, we find, for the energy density,
\begin{equation}
 w   =3\mu g^2\frac{1+\nu}{1-2\nu}
\end{equation}
and, for the total elastic energy of an isotropically
expanded sphere, 
\begin{equation}
    E_{\rm isotropic}=
    4\pi R^3\mu g^2\frac{1+\nu}{1-2\nu}.
    \label{EQ48}
\end{equation}
In the case of an isotropically expanded
spherical void, the displacement
is ${\bf u}=g\frac{R^3}{r^2}{\bf r}$.
The corresponding energy density is
\begin{equation}
    w= \frac{6\mu g^2 R^6}{r^6}
\end{equation}
and the corresponding total energy is
\begin{equation}
    E_{\rm isotropic}= 8\pi \mu g^2 R^3.
\end{equation}

\begin{widetext}
\begin{table*}[ht]
\caption{Spherical harmonics components of shape, displacement and strain}
\centering
\begin{tabular}{c| c| c| c| c}
\hline
\hline
& $l$ (sphere) & $m$ (sphere) & $l $ (void) & $m $ (void) \\
 
 \hline
 Shape & $l$&$\pm m$&$ l$&$\pm m$ \\
 \hline
 $u_{x,y}(R), v_{x,y}(R)$&$l\pm1$&$\pm m\pm1$&$l\pm1$&$\pm m\pm1$\\
 \hline
  $u_z(R),v_z(R)$&$l\pm1$&$\pm m$&$l\pm1$&$\pm m$\\
 \hline
 $u_{x,y}, v_{x,y}$&$l\pm1$&$\pm m\pm1$&$l\pm1$&$\pm m\pm1$\\
 \hline
  $u_z, v_z$&$l\pm1$&$\pm m$&$l\pm1$&$\pm m$\\
 \hline
 $\epsilon_{xx,xy,yx,yy}$&$l,l-2$&$\pm m\pm 2,\pm m$&$l,l+2$&$\pm m\pm 2,\pm m$\\
 \hline
 $\epsilon_{xz,zx,yz,zy}$&$l,l-2$&$\pm m\pm 1$&$l,l+2$&$\pm m\pm 1$\\
 \hline
 $\epsilon_{zz}$&$l,l-2$&$\pm m$&$l,l+2$&$\pm m$\\
 \hline
 \hline
\end{tabular}
\label{Table1}
\end{table*}
\end{widetext}

\section{Core-shell system}
\label{coreshell} 
In this section, we revisit the buckling instability
that occurs in a spherical core-shell system,
when the area mismatch
between a stiff shell and a soft core
exceeds a critical value,
corresponding to the elastic
energy of a isotropically expanded state
exceeding the elastic energy of a buckled state.
%
To generally treat a core-shell system, composed of
two materials with different elastic properties,
in addition to the regular solution applicable within the core,
we would also need the solution to Eq.~\ref{eom}
within a spherical shell.
The solution within a shell
is the superposition
of the regular and the irregular solutions,
which must then together be matched to the appropriate
boundary conditions at the inner radius, where the core and the
shell meet, and at the outer
radius of the shell. With these solutions in hand, we would
then calculate the strains and elastic energies.

Instead of this route,
we follow  Ref.~\onlinecite{PhysRevE.88.052404}
and consider the limiting case that the shell
can be described as a thin membrane of
fixed area, $A$, and bending stiffness, $\kappa$. The surface energy is calculated by integrating the square of the mean curvature, $H$, over the surface:
\begin{equation}
E_{\text{surface}}=\frac{\kappa}{2}\oint_{r=R}H^2 dS .
\end{equation}
Then,
when the shape of the membrane is described by a real spherical harmonic, $Y_{lm}$,
the $l$-dependent part of the membrane
elastic energy is
    \begin{equation} 
    E_{\text{surface}}=\frac{\kappa}{8}
    g^2[l(l+2)(l^2-1)],
    \label{EQ51}
    \end{equation}
independent of $m$ \cite{PhysRevA.36.4371,PhysRevE.88.052404}.

In the context of a fixed-area membrane,
 the buckling instability is controlled by
relative excess area,
namely the difference between the area of the membrane
and the area of the spherical core, normalized by
the area of the core:
\begin{equation}
    \Delta=\frac{A}{4\pi R^2}-1.
\end{equation}
Therefore, we must relate the buckling amplitude, $g$, to
the relative excess area, $\Delta$.
For buckled shapes, described by real spherical harmonics, $Y_{lm}$, Ref.~\onlinecite{PhysRevE.88.052404} showed that
\begin{equation}
    g=\sqrt{\Delta}(\frac{8\pi}{l(l+1)+2})^{1/2}.
    \label{EQ53}
\end{equation}
In this case, the energies of both the core and the shell
are proportional to $g^2$. Therefore,
the total energy of the core-shell system with shape $Y_{lm}$ is proportional to $\Delta$.

Combining Eq.~\ref{EQ46},
Eq.~\ref{EQ51},
and Eq.~\ref{EQ53}
and introducing $\alpha$, given by
\begin{equation}
    \alpha =\frac{\mu R^3}{\kappa}
\end{equation}
we find that the total
core-shell energy for spheres is
\begin{widetext}
\begin{equation}
 E_{\rm total}=   \frac{8 \kappa \Delta}{l^2+l+2}\pi \left(\alpha \frac{\left(2 l^2-3l-1\right)\nu -\left(2l^2-l+1\right)}{2
(2 l+1)\nu -(3 l+1)}
+\frac{1}{8}(l-1) l (l+1) (l+2)
\right).\\
\end{equation}
Similarly, the
total energy for spherical voids,
with a membrane surrounding the void, is
\begin{equation}
E_{\rm total}=\frac{8\kappa \Delta}{l^2+l+2}\pi \left(\alpha \frac{\left(4+7 l+2 l^2\right)\nu -\left(4+5
l+2 l^2\right)}{2 (1+2 l)\nu -(2+3 l)}
+\frac{1}{8}(l-1) l (l+1) (l+2)
\right).
\end{equation}
\end{widetext}

Fig.~\ref{reg.energy.plot} and Fig.~\ref{irreg.energy.plot}
plot these energies for $\nu = 0.2$.
In these plots, each line represents the energy associated with a particular value of $l$.
It is clear from these figures that which value of $l$ corresponds to the lowest energy
steps from one value to the next as $\alpha$ increases.
The $l$ value of the lowest total energy state is
plotted versus $\alpha$ in Fig. \ref{lmin_r} for spheres and in Fig. \ref{lmin_i} for spherical voids.
As $\alpha$ increases, the minimum energy $l$ increases. We pick four values of Poisson's ratio $\nu$ to illustrate the trend. Poisson's ratio is the material property describing the deformation of a material in directions perpendicular to the direction of loading, which lies between $-1$ and $\frac{1}{2}$ for stable, isotropic, linear elastic material.
A Poisson's ratio of $\frac{1}{2}$ means that the material is incompressible.\par

To further make sense of Fig.~\ref{lmin_r},
we consider the limit of large $l$
and treat $l$ as a continuous variable.
Then, for spheres
\begin{equation}
    E_{\rm total} \simeq 8 \kappa \Delta
    \left ( \frac{2 \alpha (1-\nu) }{(3-4\nu)l} +\frac{l^2}{8} \right ),
\end{equation}
and
we can find the value of $l$ that minimizes the total energy ($l^*$).  The result is
\begin{equation}
    l^* = 2 \left ( \frac{\alpha(1-\nu) }{3-4\nu} \right )^\frac{1}{3} 
\end{equation}
The value of $l^*$ varies as $\alpha^\frac{1}{3}$, consistent with the behavior apparent in Fig.~\ref{lmin_r}.
The  elastic energy corresponding to $l^*$ is
\begin{equation}
    E^*_{\rm total} = 12 \kappa \Delta
    \left ( \frac{\alpha (1-\nu)}{3-4 \nu} \right )^\frac{2}{3},
\end{equation}
reminiscent of the minimum energy envelope in Fig.~\ref{reg.energy.plot}.

For the isotropically expanded state,
$g=\Delta/2$ to linear order.
Therefore,
in contrast to the linear-in-$\Delta$ buckled state energy, 
the energy of an isotropically expanded sphere is, 
\begin{equation}
    E_{\rm isotropic}=\pi R^3\mu \Delta^2\frac{1+\nu}{1-2\nu},
\end{equation}
proportional to $\Delta^2$.
Thus, for small $\Delta$, the isotropic state inevitably
has a lower energy than the buckled state, while for large $\Delta$,  the  opposite is true.

To find the critical value of $\Delta$ at which the core-shell system transitions from isotropically expanded to buckled, we set the energies for both cases to be equal,
and then solve for the corresponding value of $\Delta$,
namely $\Delta_c$:
\begin{equation}
   \Delta_{c}
    =\frac{E_{\text{total}}}{\pi R^3\mu \frac{1+\nu}{1-2\nu} \Delta_c}
    =\frac{E_{\text{total}}/(\kappa \Delta_c)}{\pi \alpha \frac{1+\nu}{1-2\nu}}.
    \label{EQ55}
\end{equation}
Since $E_{\text{total}}/(\kappa \Delta_c)$ is independent of
$\Delta_c$, the right-hand side of Eq.~\ref{EQ55} is the
desired solution for $\Delta_c$.
The analogous result for spherical voids with an
interior shell is
\begin{equation}
    \Delta_c=\frac{E_{\rm total}/(\kappa \Delta_c)}{2 \pi \alpha}.
\end{equation}
The deformations
shown in Fig.~\ref{spherexy} and
Fig.~\ref{voidxy} for spheres and spherical voids,
respectively, both
correspond to $\Delta_c$ for $\alpha=600$.
We plot $\Delta_{c}$ as a function of $\alpha$ as the curved lines in  Fig. \ref{delta_crit_r} for spheres and in Fig. \ref{delta_crit_i} for spherical voids. The region below the $\Delta_{c}$-versus-$\alpha$ curve corresponds
to an isotropically expanded phase, while the region above is the buckled phase. The vertical lines in these figures separate buckled phases with different $l$ values. Thus,
Fig. \ref{delta_crit_r}  and Fig. \ref{delta_crit_i} represent shape phase diagrams.

In general, a larger value of $\alpha$ requires a lower relative excess area in order for
there to be a transition into the buckled phase above the $\Delta_{c}$ curve.
A larger value of $\alpha$
also gives rise to a larger $l$ value in the buckled phase.
Clearly, the vertical lines separating
buckled states with different values of $l$ do not align
at the same values of $\alpha$ for different Poisson ratios.
In the large-$l$ limit, for a core-shell system,
we have that
\begin{equation}
    \Delta_c = \frac{12}{\pi} \alpha^{-\frac{1}{3}} \left ( \frac{1-2\nu}{1+\nu} \right )\left (\frac{1-\nu}{3-4\nu} \right )^\frac{2}{3}.
\end{equation}
\par

\section{Conclusion} \label{discussion}

By applying linear elasticity theory and
exploiting well-known properties of the solid harmonics, we have described how to find the
displacements
either inside solid spheres
or
outside spherical voids, assuming in both cases that the
surface of the sphere or the void shows
a radial surface deformations, whose amplitude is
given by a real spherical harmonic.
Using the displacements so-obtained, we then calculated the corresponding  bulk elastic energies, providing closed-form expressions
for these energies,
for any values of the spherical harmonic degree ($l$), Poisson ratio, and shear modulus. We found that the elastic energies are independent of the spherical harmonic index ($m$),
consistent with expectations based on symmetry considerations.
These collected results represent an
important addition to our
knowledge of the linear elasticity of systems
with (near) spherical symmetry.
In addition to their relevance to the buckling/wrinkling transitions
of core-shell systems, 
because any shape can be
described as a superposition
of spherical harmonics,
our results  will be valuable for
researchers broadly interested in the elasticity of
spheres or spherical voids, that experience
surface shape deformations.
We also revisited the buckling instability experienced by a core-shell system comprising an elastic sphere, attached within a membrane of fixed area, that occurs when the area of the membrane sufficiently exceeds the area of the unstrained sphere.
By finding the state which possesses the smallest total energy the sum of the bulk and surface elastic energies
within linear elasticity, we determined the phase diagram of the core-shell sphere's shape, specifying what value of $l$ is realized as a function of the area mismatch and the core-shell elasticity. Similarly, we also determined the shape phase diagram for a spherical void bounded by a fixed-area membrane.

\section*{Supplementary Material}
A Mathematica
notebook that performs the calculations described is
available as supplementary
material \cite{URL}.

\section*{Acknowledgements}
This work was supported by an Allen Distinguished Investigator Award, a Paul G. Allen Frontiers Group advised grant of the Paul G. Allen Family Foundation.
We are especially grateful to David Poland for finding
the simple form of the bulk energy
for general values of $l$, and Nick Read for invaluable discussions.
\appendix
\section {Properties of regular solid harmonics}
\label{appendixA}

We summarize some useful results: 
\begin{equation} \label{1}
    \nabla^2(r^lY_l^m)=0,
\end{equation}
\begin{eqnarray} \label{2}
    \nabla\cdot (r^2\nabla(r^lY_l^m)) &=(\nabla r^2)\cdot \nabla(r^lY_l^m) \nonumber \\
    &=2\mathbf{r}\cdot \nabla(r^lY_l^m) \nonumber \\
    &=2r\frac{\partial}{\partial r}(r^lY_l^m) \nonumber\\
    &=2l(r^lY_l^m),
\end{eqnarray}

\begin{equation}
    \nabla^2(r^2\nabla(r^lY_l^m))=2(2l+1)\nabla(r^lY_l^m),
\end{equation}
\begin{widetext}
\begin{equation} \label{dYdx}
    \frac{\partial}{\partial x}(r^lY_l^m)=
\frac{1}{2} r^{l-1} \sqrt{\frac{(2 l+1) (l-m)!}{(l+m)!}} \left(\frac{Y_{l-1}^{m+1}}{\sqrt{\frac{(2 l-1) (l-m-2)!}{(l+m)!}}}-\frac{(l+m-1)
(l+m) Y_{l-1}^{m-1}}{\sqrt{\frac{(2 l-1) (l-m)!}{(l+m-2)!}}}\right),
\end{equation}
\begin{equation} \label{dYdy}
    \frac{\partial}{\partial y}(r^lY_l^m)=
-\frac{1}{2} i r^{l-1} \sqrt{\frac{(2 l+1) (l-m)!}{(l+m)!}} \left(\frac{(l+m-1) (l+m) Y_{l-1}^{m-1}}{\sqrt{\frac{(2 l-1)
(l-m)!}{(l+m-2)!}}}+\frac{Y_{l-1}^{m+1}}{\sqrt{\frac{(2 l-1) (l-m-2)!}{(l+m)!}}}\right),
\end{equation}
\end{widetext}
and
\begin{equation} \label{dYdz}
    \frac{\partial}{\partial z}(r^lY_l^m)=
\frac{(l+m) r^{l-1} \sqrt{\frac{(2 l+1) (l-m)!}{(l+m)!}} Y_{l-1}^m}{\sqrt{\frac{(2 l-1) (l-m-1)!}{(l+m-1)!}}}.
\end{equation}
It follows that
\begin{equation}
    \alpha=\frac{(b_x-i b_y) \sqrt{\frac{(2 l+3) (l-m+1)!}{(l+m+1)!}}}{2 \sqrt{\frac{(2 l+1) (l-m-1)!}{(l+m+1)!}}},
\end{equation}
\begin{equation}
    \beta=\frac{b_z (l+m+1) \sqrt{\frac{(2 l+3) (l-m+1)!}{(l+m+1)!}}}{\sqrt{\frac{(2 l+1) (l-m)!}{(l+m)!}}},
\end{equation}
and
\begin{equation}
    \gamma=-\frac{(b_x+i b_y) (l+m) (l+m+1) \sqrt{\frac{(2 l+3) (l-m+1)!}{(l+m+1)!}}}{2 \sqrt{\frac{(2 l+1) (l-m+1)!}{(l+m-1)!}}}.
\end{equation}

\begin{widetext}
\section {Properties of irregular solid harmonics}
\label{appendixB}
\begin{equation}
    \frac{\partial}{\partial x}(r^{-l-1} Y_l^m)=\frac{1}{2} \sqrt{\frac{2 l+1}{2 l+3}} \left(\sqrt{(l+m+1) (l+m+2)} r^{-l-2} Y_{l+1}^{m+1}-\sqrt{(l-m+1) (l-m+2)} r^{-l-2}
Y_{l+1}^{m-1}\right)
\end{equation}

\begin{equation}
    \frac{\partial}{\partial y}(r^{-l-1} Y_l^m)=-\frac{1}{2} i \sqrt{\frac{2 l+1}{2 l+3}} \left(\sqrt{(l-m+1) (l-m+2)} r^{-l-2} Y_{l+1}^{m-1}+\sqrt{(l+m+1) (l+m+2)} r^{-l-2}
Y_{l+1}^{m+1}\right)
\end{equation}

\begin{equation}
    \frac{\partial}{\partial z}(r^{-l-1} Y_l^m)=-\sqrt{\frac{2 l+1}{2 l+3}} \sqrt{(l-m+1) (l+m+1)} r^{-l-2} Y_{l+1}^m
\end{equation}


\section{Regular solution coefficients}
\label{AppendixC}
\begin{equation}
A_{m+1}=
\left\{
\begin{array}{ll}
 0 & l<m+2 \\
 \frac{g (l+1) (2 l+3) e^{2 i \pi  (l+m)} R^{-l-1} \Gamma (l-m+1)}{4 \sqrt{2} (l (4 \nu -3)+2 \nu -1) \sqrt{\left(4 l^2-1\right) (l-m)! \Gamma (l-m-1)}}
& \text{otherwise} \\
\end{array}
\right.
\end{equation}

\begin{equation}
    A_{m-1}=
 \left\{
\begin{array}{ll}
 \frac{g (-1)^{-2 l} \sqrt{\frac{(l+m-1) (l+m)}{8 l^2-2}} R^{-l-1}  \left(-(l+1)
(2 l+3) e^{2 i \pi  (2 l+m)}   \right)}{4 (l (4 \nu -3)+2 \nu -1)  } & l+m\geq 2 \\
 0 & \text{otherwise} \\
\end{array}
\right.
\end{equation}

\begin{equation}
    B_{m+1}=
-\frac{1}{2} g (-1)^{2 (l+m)} \sqrt{\frac{(l+m+1) (l+m+2)}{8 l (l+2)+6}} R^{-l-1}
\end{equation}

\begin{equation}
    B_{m-1}=
\frac{1}{2} g (-1)^{2 (l+m)} \sqrt{\frac{(l-m+1) (l-m+2)}{8 l (l+2)+6}} R^{-l-1}
\end{equation}

\begin{equation}
    C_{m+1}=
 \left\{
\begin{array}{ll}
 \frac{1}{2} g (-1)^{-2 l} \sqrt{\frac{(l-m-1) (l-m)}{8 l^2-2}} R^{l+1} \left(R^2\right)^{-l} & l\geq m+2 \\
 0 & \text{otherwise} \\
\end{array}
\right.
\end{equation}

\begin{equation}
    C_{m-1}=
 \left\{
\begin{array}{ll}
 -\frac{1}{2} g (-1)^{-2 l} \sqrt{\frac{(l+m-1) (l+m)}{8 l^2-2}} R^{l+1} \left(R^2\right)^{-l} & l+m\geq 2 \\
 0 & \text{otherwise} \\
\end{array}
\right.
\end{equation}

\begin{equation}
    D_{m+1}=\begin{cases}
    \!\begin{aligned}
&\frac{1}{8 (2 l+1)^2 (l (4 \nu -3)+2 \nu -1) \sqrt{l (8 l (l+1)-6) \Gamma (l-m-1)}}(i g (-1)^m e^{i \pi  (3 l+m)} (-R)^{-l-1} \\&(l^2 \sqrt{l (2 l+1)^3 (2 l+3) \Gamma (l-m+1)}+l (2 m \sqrt{l (2 l+1)^3 (2 l+3) \Gamma (l-m+1)}\\&+3 \sqrt{l (2 l+1)^3 (2 l+3) \Gamma (l-m+1)}+2 \sqrt{l (2 l+1)^3 (2 l+3) (l-m+1) \Gamma (l-m+2)}\\&+2 \sqrt{l (2 l+1) (2 l+3) (l-m+1) (l-m+2) \Gamma (l-m+3)})+2 \sqrt{l (2 l+1)^3 (2 l+3) \Gamma (l-m+1)}\\&+2 \sqrt{l (2 l+1)^3 (2 l+3) (l-m+1) \Gamma (l-m+2)}+m (m \sqrt{l (2 l+1)^3 (2 l+3) \Gamma (l-m+1)}\\&+3 \sqrt{l (2 l+1)^3 (2 l+3) \Gamma (l-m+1)}+2 \sqrt{l (2 l+1)^3 (2 l+3) (l-m+1) \Gamma (l-m+2)})\\&+\sqrt{l (2 l+1) (2 l+3) (l-m+1) (l-m+2) \Gamma (l-m+3)}))
    \end{aligned}
& l\geq m+2 \\
 0 & \text{otherwise} \\
\end{cases}
\end{equation}

\begin{equation}
    D_{m-1}=\begin{cases}
    \!\begin{aligned}
\frac{i g (l+1) (2 l+3) e^{i \pi  (3 l+2 m)} \sqrt{\frac{(l+m-1) (l+m)}{8 l^2-2}} (-R)^{-l-1}}{4 (l (4 \nu -3)+2 \nu -1)}
    \end{aligned}
& l+m\geq 2 \\
 0 & \text{otherwise} \\
\end{cases}
\end{equation}

\begin{equation}
    E_{m+1}=
\frac{1}{2} i g (-1)^{2 (l+m)} \sqrt{\frac{(l+m+1) (l+m+2)}{8 l (l+2)+6}} R^{-l-1}
\end{equation}

\begin{equation}
    E_{m-1}=
\frac{1}{2} i g (-1)^{2 (l+m)} \sqrt{\frac{(l-m+1) (l-m+2)}{8 l (l+2)+6}} R^{-l-1}
\end{equation}

\begin{equation}
    F_{m+1}=
 \left\{
\begin{array}{ll}
 -\frac{1}{2} i g (-1)^{-2 l} \sqrt{\frac{(l-m-1) (l-m)}{8 l^2-2}} R^{l+1} \left(R^2\right)^{-l} & l\geq m+2 \\
 0 & \text{otherwise} \\
\end{array}
\right.
\end{equation}

\begin{equation}
    F_{m-1}=
 \left\{
\begin{array}{ll}
 -\frac{1}{2} i g (-1)^{-2 l} \sqrt{\frac{(l+m-1) (l+m)}{8 l^2-2}} R^{l+1} \left(R^2\right)^{-l} & l+m\geq 2 \\
 0 & \text{otherwise} \\
\end{array}
\right.
\end{equation}

\begin{equation}
    I=
 \left\{
\begin{array}{ll}
 g (-1)^{-2 l} \sqrt{\frac{(l-m) (l+m)}{8 l^2-2}} R^{l+1} \left(R^2\right)^{-l} & l\geq m+1 \\
 0 & \text{otherwise} \\
\end{array}
\right.
\end{equation}

\begin{equation}
    H=
g (-1)^{2 (l+m)} \sqrt{\frac{(l-m+1) (l+m+1)}{8 l (l+2)+6}} R^{-l-1}
\end{equation}

\begin{equation}
    G=\begin{cases}
\begin{split}
 &\frac{1}{4 \sqrt{2} (l (4 \nu
-3)+2 \nu -1) \sqrt{(2 l-1) \Gamma (l-m) \Gamma (l+m)}}(g (-1)^{2 (l+m)} R^{-l-1} \\&(((l+m+1) (l+m+2))^{3/2} \Gamma (l+m+1) \sqrt{\frac{\Gamma (l-m+1)}{(2 l+1) \Gamma (l+m+3)}}\\&+\sqrt{\frac{(l-m+1)
(l-m+2) \Gamma (l-m+3) \Gamma (l+m+1)}{2 l+1}}\\&+2 \sqrt{\frac{(l-m+1) (l+m+1) \Gamma (l-m+2) \Gamma (l+m+2)}{2 l+1}}))
\end{split}& l\geq m+1 \\
 0 & \text{otherwise} \\
\end{cases}
\end{equation}

\section{Irregular solution coefficients}
\label{AppendixD}
\begin{equation}
    \begin{split}
J_{m+1}=&\frac{1}{8 (4 l \nu -3 l+2 \nu -2)}g (-1)^{m-l} \sqrt{\frac{(l+m+1) (l+m+2)}{4 l (l+2)+3}} R^l (\sqrt{l \left(4 l^2-1\right) (l-m-1)
(l-m)} \\&\left(
\left\{
\begin{array}{ll}
 (-1)^{-l-m} \sqrt{\frac{(l-m-1) (l-m)}{8 l^3-2 l}} & l\geq m+2 \\
 0 & \text{otherwise} \\
\end{array}
\right.
\right)\\&+\sqrt{2} \sqrt{l \left(4 l^2-1\right) (l-m) (l+m)} \left(
 \left\{
\begin{array}{ll}
 (-1)^{-l-m} \sqrt{-\frac{(l-m) (l+m)}{l-4 l^3}} & l\geq m+1 \\
 0 & \text{otherwise} \\
\end{array}
\right.
\right)\\&+\sqrt{l \left(4 l^2-1\right) (l+m-1) (l+m)} \left(
\left\{
\begin{array}{ll}
 (-1)^{-l-m} \sqrt{\frac{(l+m-1) (l+m)}{8 l^3-2 l}} & l+m\geq 2 \\
 0 & \text{otherwise} \\
\end{array}
\right.
\right))
\end{split}
\end{equation}

\begin{equation}
    J_{m-1}=
 \left\{
\begin{array}{ll}
 \frac{g (-1)^{-2 l} (1-2 l) l \sqrt{\frac{(l-m+1) (l-m+2)}{8 l (l+2)+6}} R^l}{4 (4 l \nu -3 l+2 \nu -2)} & l\geq m+2 \\
 -\frac{g (-1)^{-2 l} \sqrt{\frac{(l-m+1) (l-m+2)}{8 l (l+2)+6}} (3 l-m-1) (l+m) R^l}{8 (4 l \nu -3 l+2 \nu -2)} & l\geq m+1\land l+m\geq 2 \\
 -\frac{g (-1)^{-2 l} (l-m) \sqrt{\frac{(l-m+1) (l-m+2)}{8 l (l+2)+6}} (l+m) R^l}{4 (4 l \nu -3 l+2 \nu -2)} & l\geq m+1 \\
 -\frac{g (-1)^{-2 l} \sqrt{\frac{(l-m+1) (l-m+2)}{8 l (l+2)+6}} (l+m-1) (l+m) R^l}{8 (4 l \nu -3 l+2 \nu -2)} & l+m\geq 2 \\
\end{array}
\right.
\end{equation}

\begin{equation}
    K_{m+1}=
 \left\{
\begin{array}{ll}
 \frac{1}{2} g (-1)^{-2 l} \sqrt{\frac{(l-m-1) (l-m)}{8 l^2-2}} R^l & l\geq m+2 \\
 0 & \text{otherwise} \\
\end{array}
\right.
\end{equation}

\begin{equation}
    K_{m-1}=
\left\{
\begin{array}{ll}
 -\frac{1}{2} g (-1)^{-2 l} \sqrt{\frac{(l+m-1) (l+m)}{8 l^2-2}} R^l & l+m\geq 2 \\
 0 & \text{otherwise} \\
\end{array}
\right.
\end{equation}

\begin{equation}
    L_{m+1}=
-\frac{1}{2} g (-1)^{2 (l+m)} \sqrt{\frac{(l+m+1) (l+m+2)}{8 l (l+2)+6}} R^{l+2}
\end{equation}

\begin{equation}
    L_{m-1}=
\frac{g (-1)^{2 (l+m)} R^{l+2}}{2 \sqrt{\frac{8 l (l+2)+6}{(l-m+1) (l-m+2)}}}
\end{equation}

\begin{equation}
\begin{split}
M_{m+1}=
&-\frac{1}{8 (4 l \nu -3 l+2 \nu -2)}i g (-1)^{m-l} \sqrt{\frac{(l+m+1) (l+m+2)}{4 l (l+2)+3}} R^l (\sqrt{l \left(4 l^2-1\right) (l-m-1)
(l-m)} \\ &\left(
 \left\{
\begin{array}{ll}
 (-1)^{-l-m} \sqrt{\frac{(l-m-1) (l-m)}{8 l^3-2 l}} & l\geq m+2 \\
 0 & \text{otherwise} \\
\end{array}
\right.
\right)\\ &+\sqrt{2} \sqrt{l \left(4 l^2-1\right) (l-m) (l+m)} \left(
 \left\{
\begin{array}{ll}
 (-1)^{-l-m} \sqrt{-\frac{(l-m) (l+m)}{l-4 l^3}} & l\geq m+1 \\
 0 & \text{otherwise} \\
\end{array}
\right.
\right)\\&+\sqrt{l \left(4 l^2-1\right) (l+m-1) (l+m)} \left(
\left\{
\begin{array}{ll}
 (-1)^{-l-m} \sqrt{\frac{(l+m-1) (l+m)}{8 l^3-2 l}} & l+m\geq 2 \\
 0 & \text{otherwise} \\
\end{array}
\right.
\right))
\end{split}
\end{equation}

\begin{equation}
    \begin{split}
    M_{m-1}=
&-\frac{1}{8 (4 l \nu -3 l+2 \nu -2)}i g (-1)^{m-l} R^l (\sqrt{\frac{l (2 l-1) (l-m-1) (l-m) (l-m+1) (l-m+2)}{2 l+3}} \\&\left(
 \left\{
\begin{array}{ll}
 (-1)^{-l-m} \sqrt{\frac{(l-m-1) (l-m)}{8 l^3-2 l}} & l\geq m+2 \\
 0 & \text{otherwise} \\
\end{array}
\right.
\right)\\&+\sqrt{2} \sqrt{\frac{l (2 l-1) (l-m) (l-m+1) (l-m+2) (l+m)}{2 l+3}} \left(
\left\{
\begin{array}{ll}
 (-1)^{-l-m} \sqrt{-\frac{(l-m) (l+m)}{l-4 l^3}} & l\geq m+1 \\
 0 & \text{otherwise} \\
\end{array}
\right.
\right)\\&+\sqrt{\frac{l (2 l-1) (l-m+1) (l-m+2) (l+m-1) (l+m)}{2 l+3}} \left(
 \left\{
\begin{array}{ll}
 (-1)^{-l-m} \sqrt{\frac{(l+m-1) (l+m)}{8 l^3-2 l}} & l+m\geq 2 \\
 0 & \text{otherwise} \\
\end{array}
\right.
\right))
\end{split}
\end{equation}

\begin{equation}
    N_{m+1}=
 \left\{
\begin{array}{ll}
 -\frac{1}{2} i g (-1)^{-2 l} \sqrt{\frac{(l-m-1) (l-m)}{8 l^2-2}} R^l & l\geq m+2 \\
 0 & \text{otherwise} \\
\end{array}
\right.
\end{equation}

\begin{equation}
    N_{m-1}=
 \left\{
\begin{array}{ll}
 -\frac{1}{2} i g (-1)^{-2 l} \sqrt{\frac{(l+m-1) (l+m)}{8 l^2-2}} R^l & l+m\geq 2 \\
 0 & \text{otherwise} \\
\end{array}
\right.
\end{equation}

\begin{equation}
    O_{m+1}=
\frac{1}{2} i g (-1)^{2 (l+m)} \sqrt{\frac{(l+m+1) (l+m+2)}{8 l (l+2)+6}} R^{l+2}
\end{equation}

\begin{equation}
    O_{m-1}=
\frac{i g (-1)^{2 (l+m)} R^{l+2}}{2 \sqrt{\frac{8 l (l+2)+6}{(l-m+1) (l-m+2)}}}
\end{equation}

\begin{equation}
    S=
g (-1)^{2 (l+m)} \sqrt{\frac{(l-m+1) (l+m+1)}{8 l (l+2)+6}} R^{l+2}
\end{equation}

\begin{equation}
    Q=
\left\{
\begin{array}{ll}
 g (-1)^{-2 l} \sqrt{\frac{(l-m) (l+m)}{8 l^2-2}} R^l & l\geq m+1 \\
 0 & \text{otherwise} \\
\end{array}
\right.
\end{equation}

\begin{equation}
    P=
 \left\{
\begin{array}{ll}
 -\frac{g (-1)^{-2 l} l (2 l-1) \sqrt{\frac{(l-m+1) (l+m+1)}{8 l (l+2)+6}} R^l}{2 (4 l \nu -3 l+2 \nu -2)} & l\geq m+2 \\
 -\frac{g (-1)^{-2 l} (3 l-m-1) (l+m) \sqrt{\frac{(l-m+1) (l+m+1)}{8 l (l+2)+6}} R^l}{4 (4 l \nu -3 l+2 \nu -2)} & l\geq m+1\land l+m\geq 2 \\
 -\frac{g (-1)^{-2 l} (l-m) (l+m) \sqrt{\frac{(l-m+1) (l+m+1)}{8 l (l+2)+6}} R^l}{2 (4 l \nu -3 l+2 \nu -2)} & l\geq m+1 \\
 -\frac{g (-1)^{-2 l} (l+m-1) (l+m) \sqrt{\frac{(l-m+1) (l+m+1)}{8 l (l+2)+6}} R^l}{4 (4 l \nu -3 l+2 \nu -2)} & l+m\geq 2 \\
\end{array}
\right.
\end{equation}

\end{widetext}

        \begin{figure}[h]
    \centering
    \includegraphics[width=.5\textwidth]{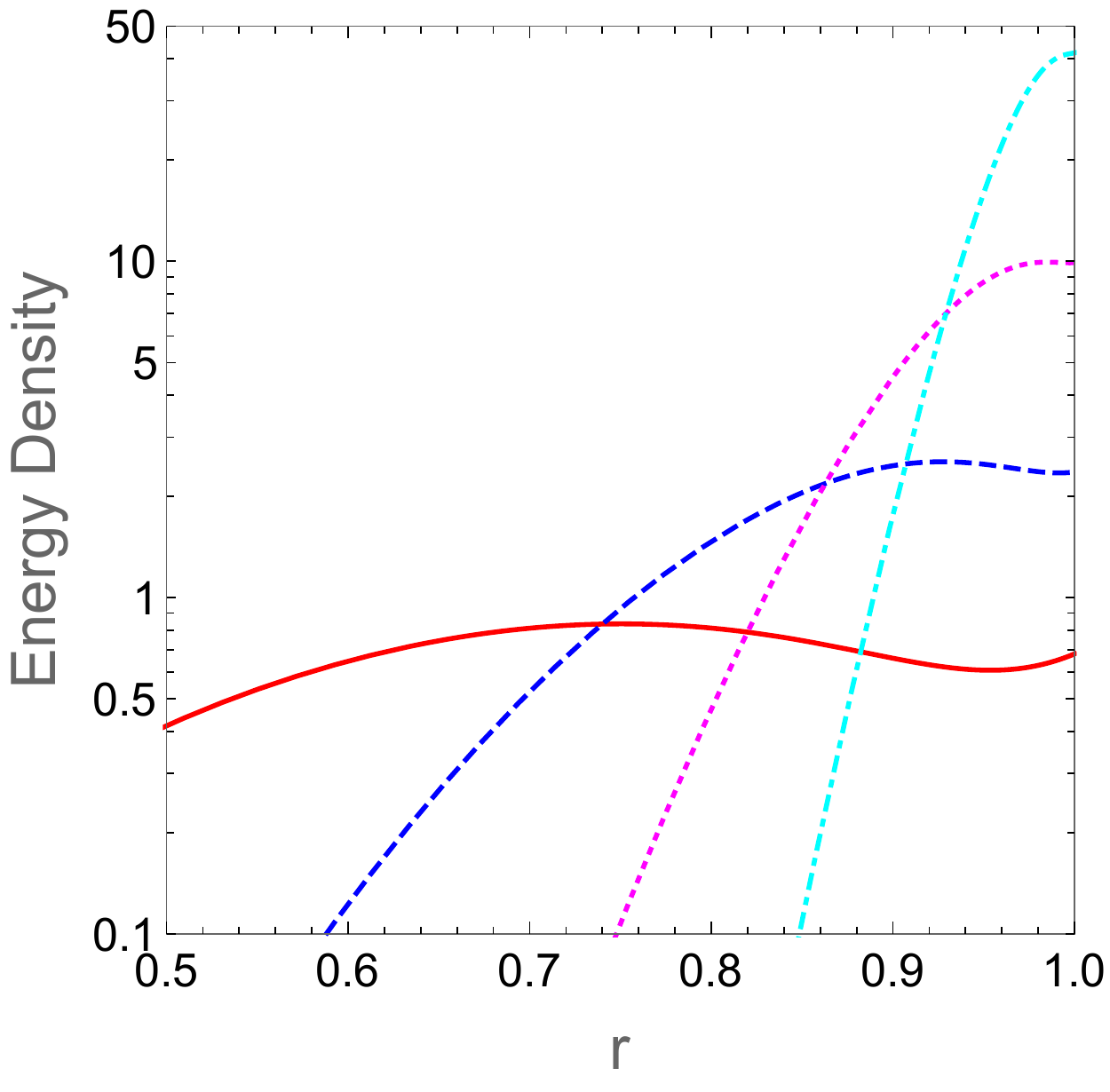}
    \caption{Plot of the energy density as a function of $r$ for $l=4,8,16,32$ for spheres. Red solid line corresponds to $l=4$, blue dashed line corresponds to $l=8$, magenta dotted line corresponds to $l=16$, and cyan dot-dashed line corresponds to $l=32$.}
    \label{energy_density_r}
\end{figure}

\begin{figure}[h]
    \centering
    \includegraphics[width=.5\textwidth]{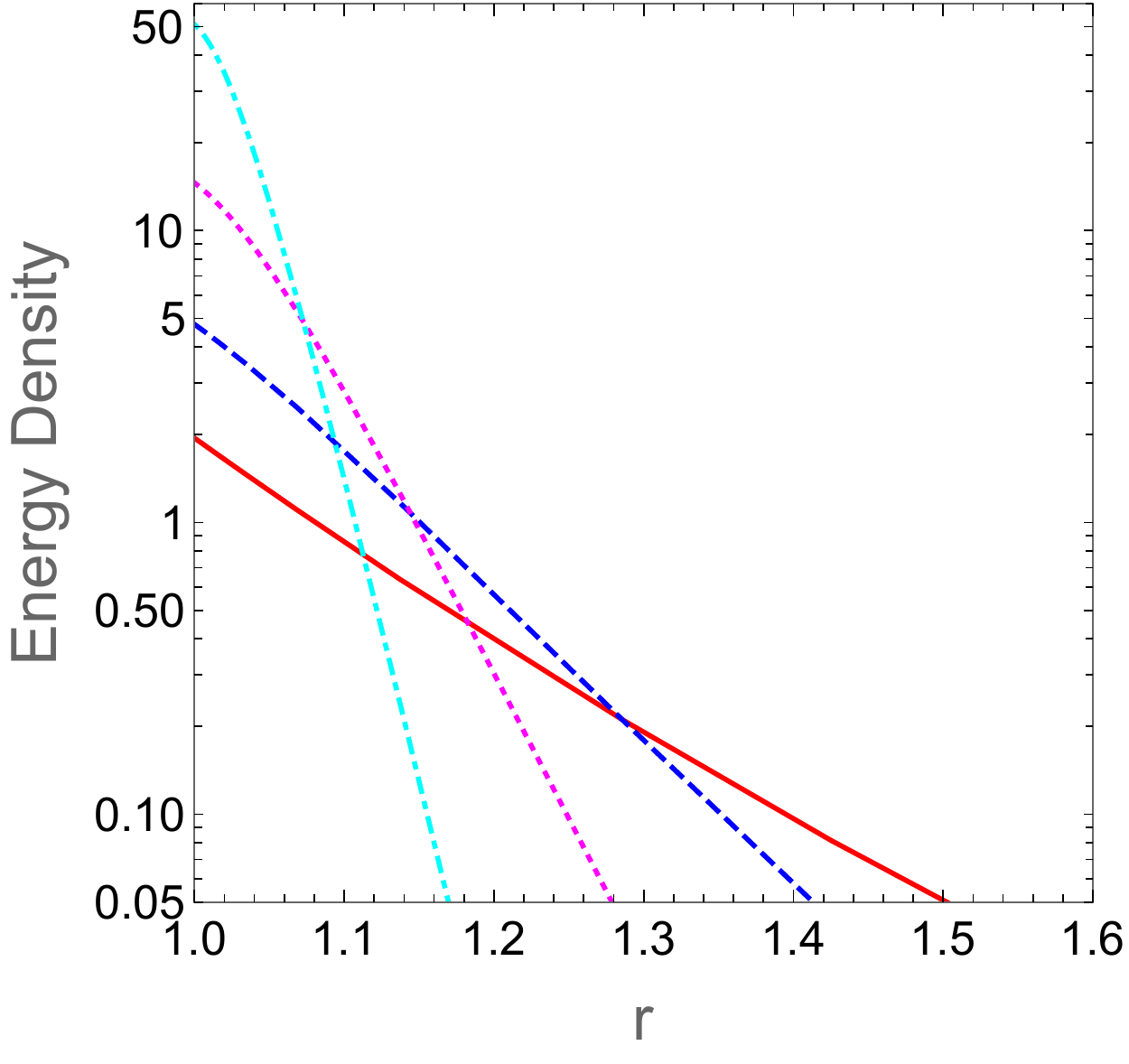}
    \caption{Plot of the energy density as a function of $r$ for $l=4,8,16,32$ for spherical voids. Red solid line corresponds to $l=4$, blue dashed line corresponds to $l=8$, magenta dotted line corresponds to $l=16$, and cyan dot-dashed line corresponds to $l=32$.}
    \label{energy_density_i}
\end{figure}

\begin{figure}[h] 
    \centering
    \includegraphics[width=.5\textwidth]{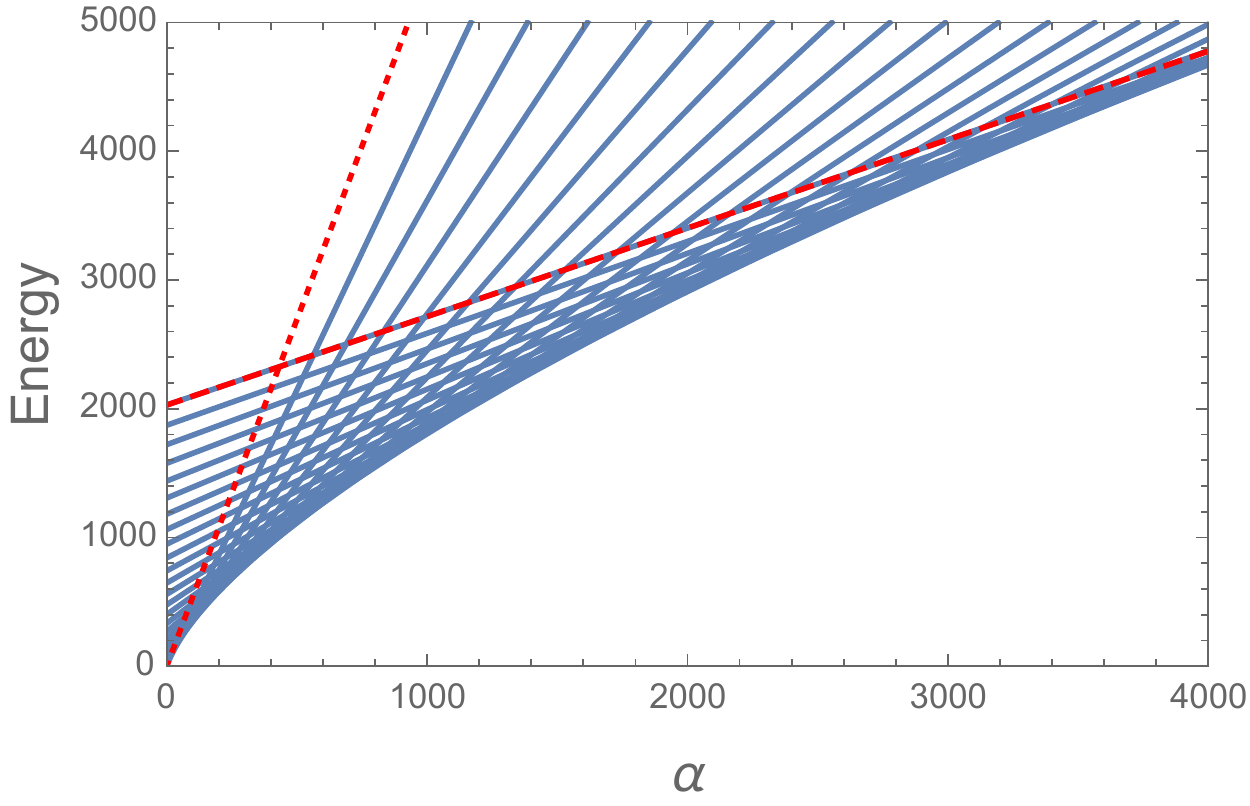}
    \caption{Plot of energy as a function of $\alpha$ for different $l$ values for the solution for spheres. Red dotted and dashed lines correspond to $l=1$ and $l=25$. Other lines correspond to $l=2,3,4,...,24$ in order of increasing distance from the red dotted line for $l=1$. $\nu = 0.2$}
    \label{reg.energy.plot}
\end{figure}

\begin{figure}[h] 
    \centering
    \includegraphics[width=.5\textwidth]{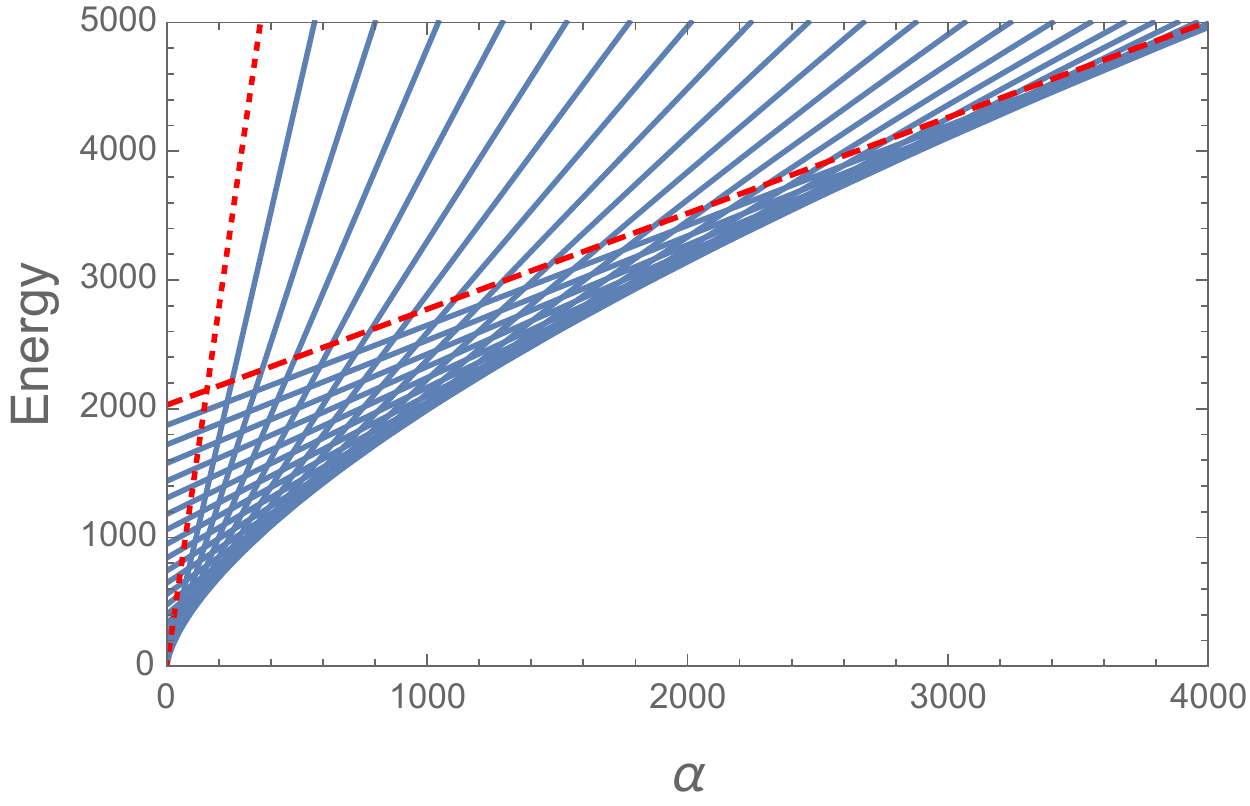}
    \caption{Plot of energy as a function of $\alpha$ for different $l$ values for the solution for spherical voids. Red dotted and dashed lines correspond to $l=1$ and $l=25$. Other lines correspond to $l=2,3,4,...,24$ in order of increasing distance from the red dotted line for $l=1$. $\nu = 0.2$}
    \label{irreg.energy.plot}
\end{figure}

\begin{figure}[h]
    \centering
    \includegraphics[width=.5\textwidth]{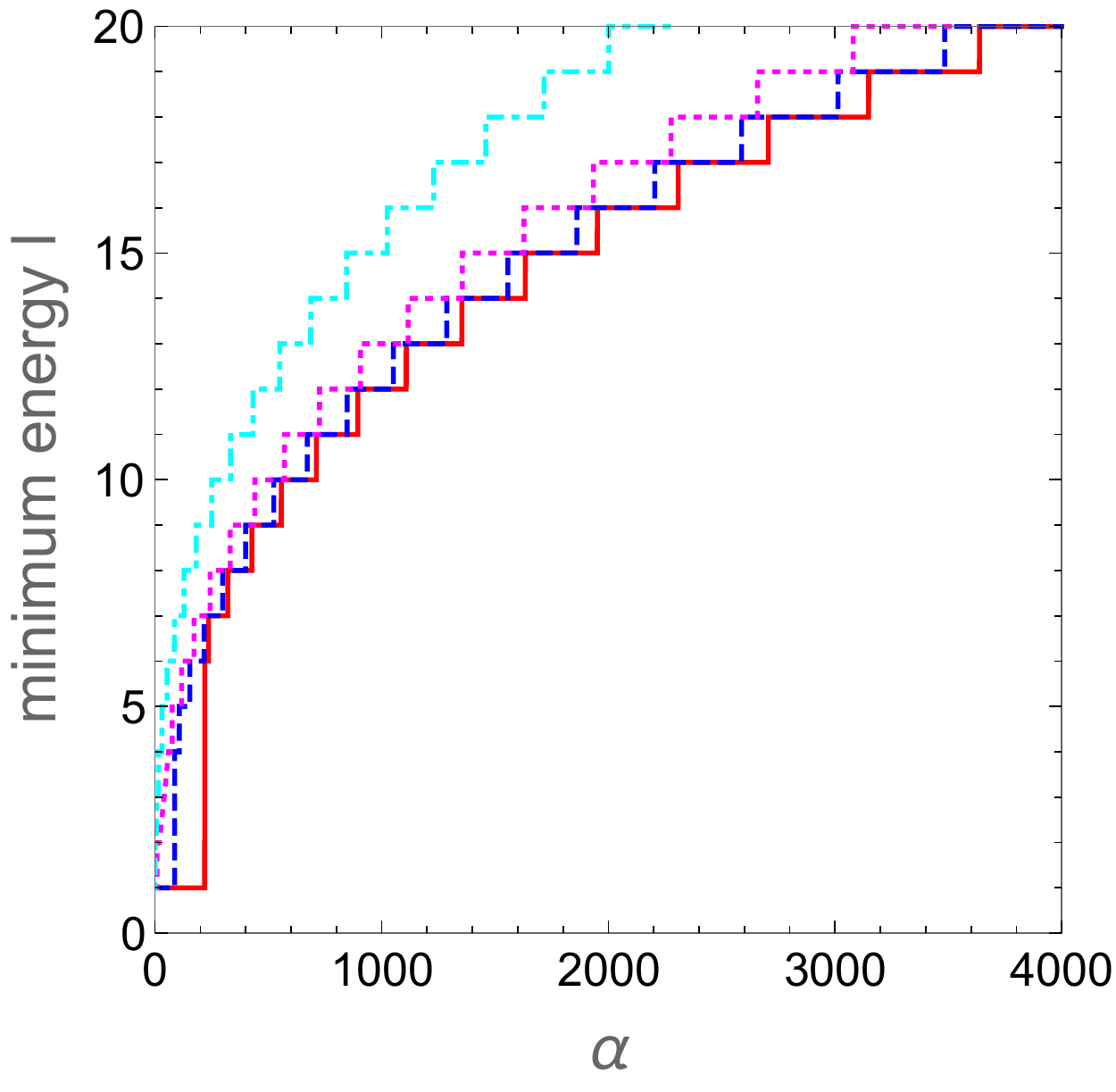}
    \caption{Plot of $l$ of the lowest energy state as a function of $\alpha$ for four different values of $\nu =-0.1,0,0.2,0.5$ for the solution for spheres. Red solid line corresponds to $\nu=-0.1$, blue dashed line corresponds to $\nu=0$, magenta dotted line corresponds to $\nu=0.2$, and cyan dot-dashed line corresponds to $\nu=0.5$.  }
    \label{lmin_r}
\end{figure}

\begin{figure}[h]
    \centering
    \includegraphics[width=.5\textwidth]{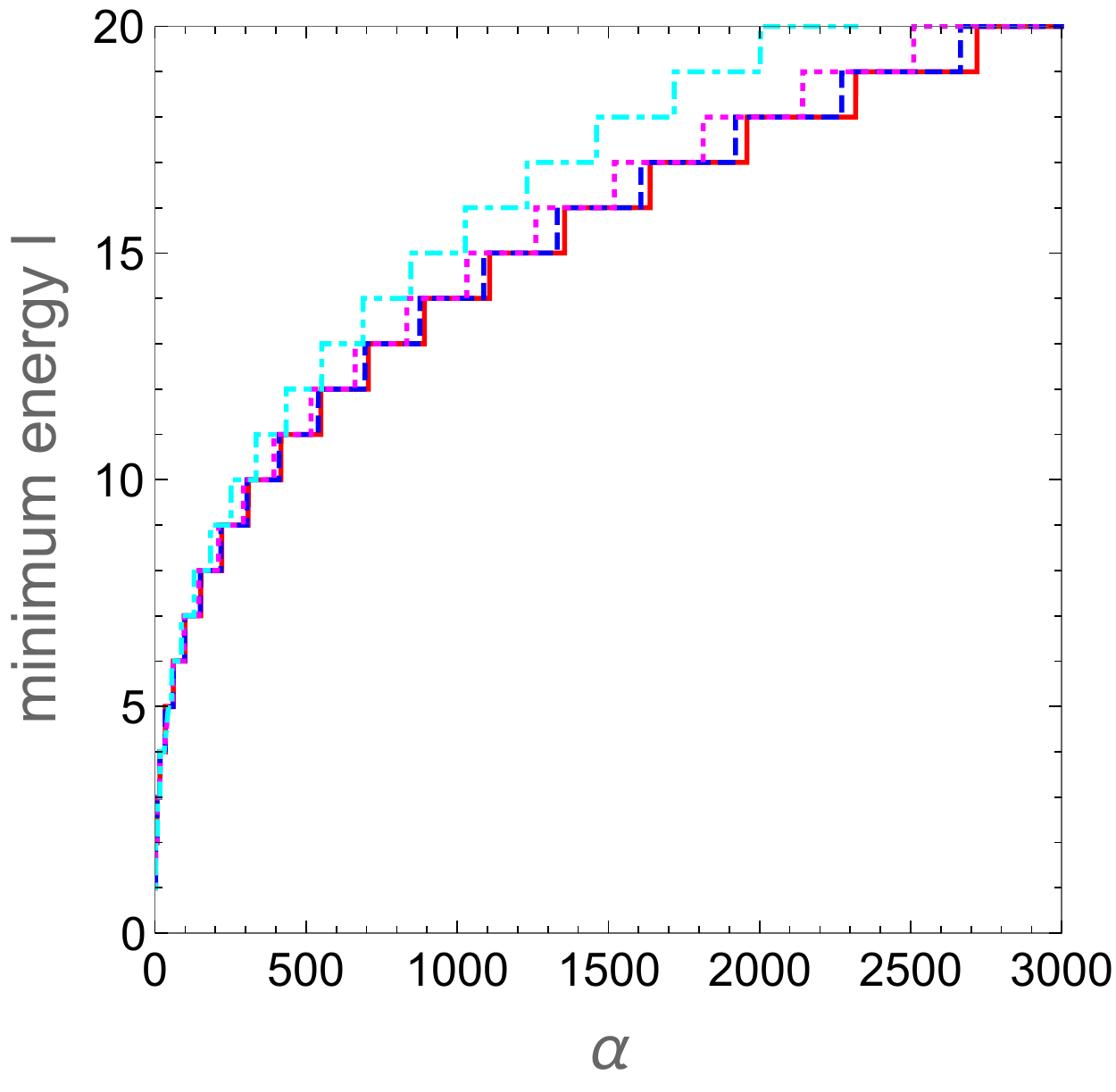}
    \caption{Plot of $l$ of the lowest energy state as a function of $\alpha$ for four different values of $\nu =-0.1,0,0.2,0.5$ for the solution for spherical voids. Red solid line corresponds to $\nu=-0.1$, blue dashed line corresponds to $\nu=0$, magenta dotted line corresponds to $\nu=0.2$, and cyan dot-dashed line corresponds to $\nu=0.5$.}
    \label{lmin_i}
\end{figure}

\begin{figure}[h]
    \centering
    \includegraphics[width=.5\textwidth]{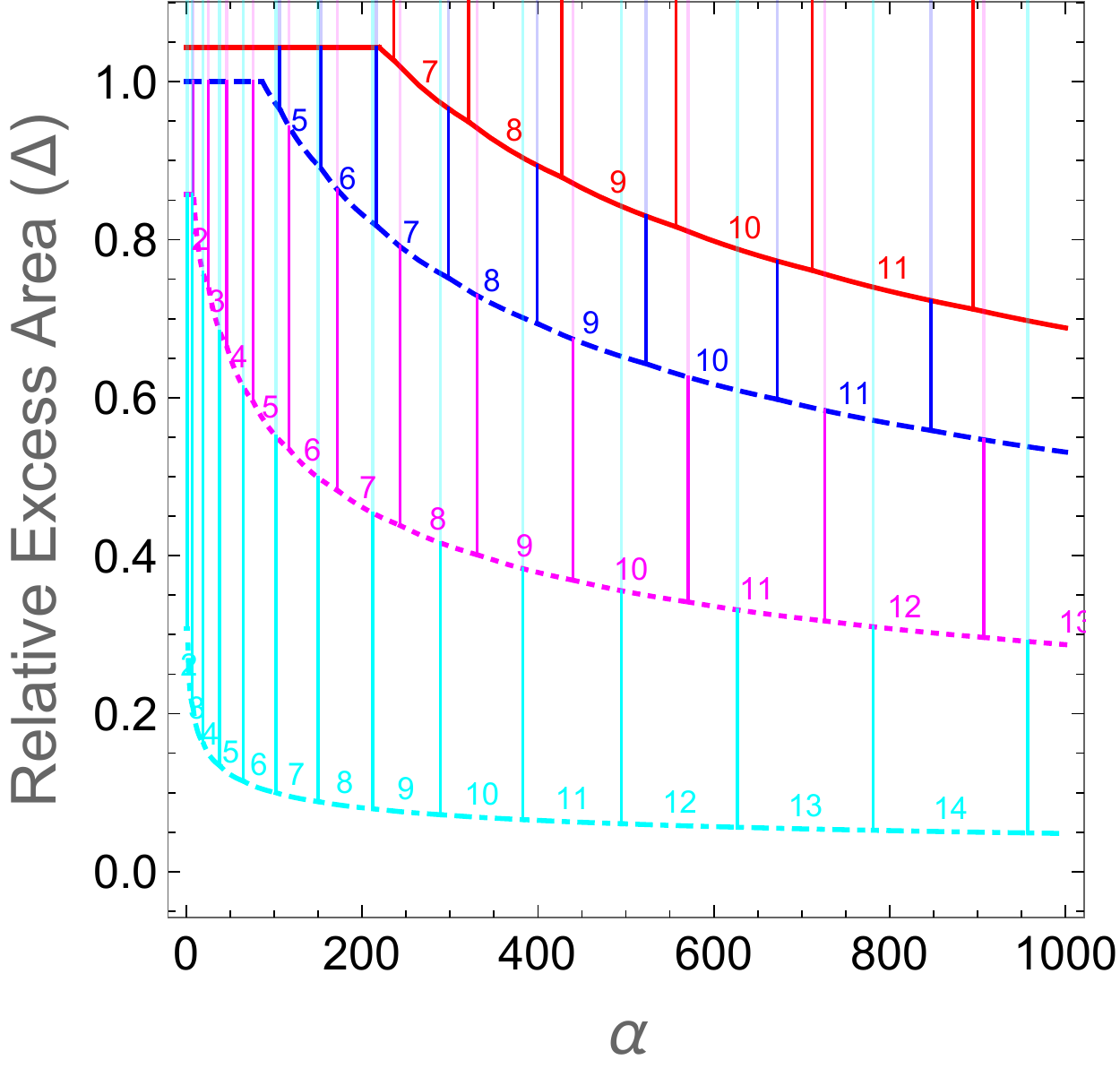}
    \caption{Sphere phase diagram for four different values of $\nu =-0.1,0,0.2,0.45$. The red solid line corresponds to $\nu=-0.1$, the blue dashed line corresponds to $\nu=0$, the magenta dotted line corresponds to $\nu=0.2$, and the cyan dot-dashed line corresponds to $\nu=0.45$. In each case, the region above these curved lines is the spherical-harmonic phase and the region below the curve is the isotropic-expansion phase. The vertical lines separate regions with different
    values of $l$.}
    \label{delta_crit_r}
\end{figure}

\begin{figure}[h]
    \centering
       \includegraphics[width=.5\textwidth]{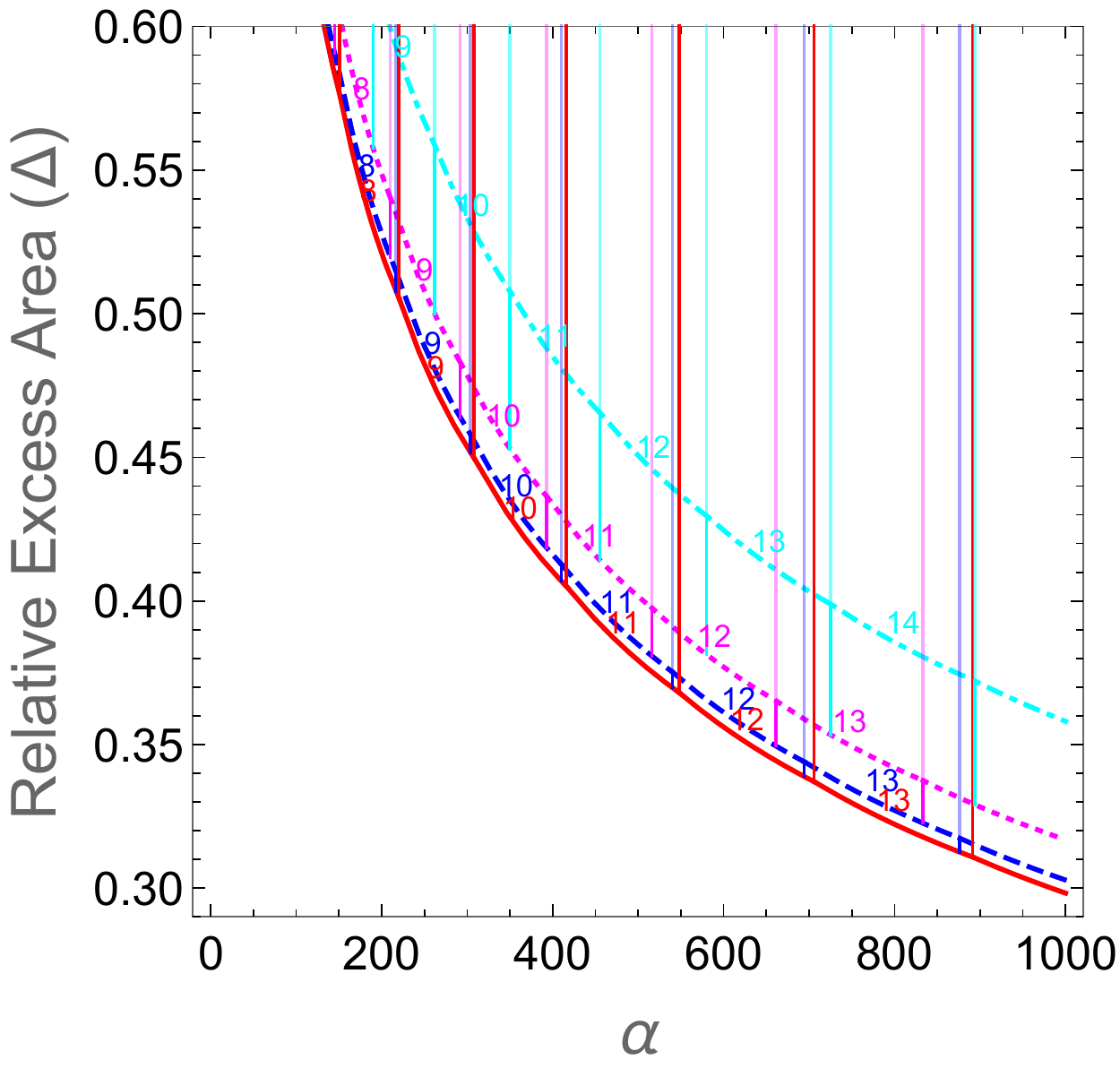}
    \caption{Spherical void phase diagram for four different values of $\nu =-0.1,0,0.2,0.45$.  The red solid line corresponds to $\nu=-0.1$, the blue dashed line corresponds to $\nu=0$, the magenta dotted line corresponds to $\nu=0.2$, and the cyan dot-dashed line corresponds to $\nu=0.45$. In each case, the region above these curved lines is the spherical-harmonic phase and the region below the curve is the isotropic-expansion phase. The vertical lines separate regions with different
    values of $l$.}
    \label{delta_crit_i}
\end{figure}


\end{document}